# Partitioning of Diluted Anyons Reveals their Braiding Statistics

June-Young M. Lee[1*], Changki Hong[2*], Tomer Alkalay[2*], Noam Schiller[3], Vladimir Umansky[2], Moty Heiblum[2], Yuval Oreg[3], H.-S. Sim[1]

[1]Department of Physics, Korea Advanced Institute of Science and Technology, Daejeon 34141, South Korea

[2]Braun Center for Submicron Research, Department of Condensed Matter Physics, Weizmann Institute of Science, Rehovot 7610001, Israel

[3]Department of Condensed Matter Physics, Weizmann Institute of Science, Rehovot 7610001, Israel

[*]Equal Contribution

Correspondent Authors:
H.-S. Sim……hs_sim@kaist.ac.kr
Moty Heiblum….moty.heiblum@weizmann.ac.il

## ABSTRACT

**Correlations of partitioned particles carry essential information about their quantumness [1]. Partitioning *full* beams of charged particles leads to current fluctuations, with their autocorrelation (namely, shot noise) revealing the particles' charge [2, 3]. This is not the case when a highly diluted beam is partitioned. Bosons or fermions will exhibit particle antibunching (due to their sparsity and discreteness) [4-6]. However, when diluted anyons, such as quasiparticles in fractional quantum Hall states, are partitioned in a narrow constriction, their autocorrelation reveals an essential aspect of their quantum exchange statistics: their braiding phase [7]. Here, we describe detailed measurements of weakly partitioned, highly diluted, one-dimension-like edge modes of the one-third filling fractional quantum Hall state. The measured autocorrelation agrees with our theory of braiding anyons in the time-domain (instead of braiding in space); with a braiding phase $2\theta=2\pi/3$, without any fitting parameters. Our work offers a relatively straightforward and simple method to observe the braiding statistics of exotic anyonic states, such as non-abelian states [8], without resorting to complex interference experiments [9].**

## Main

Fractional quantum Hall (FQH) systems host exotic quasiparticles (QPs), named anyons, that carry fractional charges and obey fractional statistics. An adiabatic braiding of abelian anyons leads to an added fractional statistical phase $2\theta$, whereas for non-abelian anyons, the original state transforms into another degenerate state [8, 10, 11]. The charge of the QPs can be determined by partitioning a *full* beam of QPs, leading to excess shot noise (autocorrelation of charge fluctuations) [2, 3]. Here, we demonstrate that partitioning a dilute anyon beam reveals the braiding phase of the QPs in the autocorrelation's Fano factor.

The traditional strategy to observe the statistics of QPs of FQH states involves interference in a Fabry-Pérot Interferometer [12, 13] or a Mach-Zehnder Interferometer [14], where edge modes circulate localized QPs in the insulating bulk. Another recent approach [15] exploited a configuration of three quantum point contacts (QPCs) where two highly dilute beams, partitioned by two side QPCs, 'collided' at a central QPC (a typical Hong-Ou-Mandel configuration [16-18]). Measured for the anyonic one-third filling FQH state, the cross-correlation of the back-scattered QPs beams was interpreted as a partly anionic bunching at the central QPC [15, 19].

A different origin of the three-QPC outcome is based on time-domain braiding between the two impinging dilute anyon beams and the thermally (or vacuum) excited particle-hole anyon pairs at the central QPC [7, 9]. To test this scenario, we focused on a two-QPC geometry where one QPC dilutes an anyon beam, further partitioned by a second QPC, resulting in excess shot noise (autocorrelation). Testing under different conditions, such as beam dilution, the second QPC's transmission, and beam travel distance, we found an anomalous autocorrelation Fano factor ($\mathcal{F}_{\text{dilute}}$) that agrees with our theory of time-domain braiding at the second (partitioning) QPC (without any fitting parameters).

Notably, although the theoretical description of the time-domain anyon braiding in a QPC is based on the chiral Luttinger liquid (CLL) theory (or the equivalent conformal field theory) [7, 9], the saddle potential in the QPCs [20] is far from the ideal barrier in the CLL theory. To overcome this difficulty, we developed a theoretical description that hybridizes the CLL theory and a phenomenological theory in the spirit of the successful ubiquitous approach of charge determination via autocorrelation measurements [2, 3].

## Shot Noise of Full Beam

Our experimental setup is shown in Fig. 1(a) (Supplementary Note I). The source (S) is biased by voltage $V_{\text{S}}$, injecting a *full* QP beam with current $I_{\text{S}} = GV_{\text{S}}$, flowing chirally along Edge1, with conductance $G = \nu e^2/h$ at filling factor $\nu = 1/3$, where $e$ is the electron charge and $h$ is the Planck constant. The full beam is highly diluted by QPC1, with a reflection probability $R_{\text{QPC1}}$ and thus current $I_{\text{QPC1}} = I_{\text{S}} R_{\text{QPC1}}$. The dilute beam flows chirally along Edge2, impinging at QPC2 (being 2 μm away), where it is further partitioned. The scattered current fluctuations are measured after being amplified by amplifiers A and B, with the spectral densities $S_{\text{A}}$, $S_{\text{B}}$ and $S_{\text{AB}}$ measured. The charge of the diluted QPs $e^*$ was determined from the autocorrelation shot noise of QPC1 [2, 3, 21-23]

$$S_{\text{QPC1}} = 2e^* I_{\text{S}} R_{\text{QPC1}} (1 - R_{\text{QPC1}}) \left[\coth\left(\frac{e^* V_{\text{S}}}{2k_{\text{B}}T}\right) - \frac{2k_{\text{B}}T}{e^* V_{\text{S}}}\right] , \tag{1}$$

which was determined by $S_{\text{QPC1}} = S_{\text{A}} + S_{\text{B}} + 2S_{\text{AB}}$, which the electron temperature $T$ and the Boltzmann constant $k_{\text{B}}$ (Fig. 1(b) and Methods). The data agrees well with Eq.(1) with $e^* = e/3$ (a similar measurement was performed with QPC2 (Supplementary Note II)).

We now elaborate on the phenomenological hybridization of the non-interacting expression in Eq. (1) and the interacting theory of the CLL. In the limit of very large $V_\mathrm{S}/T$ and very small $R_\mathrm{QPC1}$, Eq. (1) agrees with the prediction of the CLL theory. In the CLL theory, the current and shot noise are expressed as $I_\mathrm{QPC1} = e^*(W_{1\to2} - W_{2\to1})$ and $S_\mathrm{QPC1} = 2e^{*2}(W_{1\to2} + W_{2\to1})$, where $W_{i\to j}$ is the tunnelling rate of an anyon from Edge***i*** to Edge***j***. When a full (undiluted) biased beam obeys $e^*V_\mathrm{S} \gg k_\mathrm{B}T$, the rate $W_{2\to1}$ is exponentially suppressed compared with $W_{1\to2}$, resulting in $S_\mathrm{QPC1} = 2e^*I_\mathrm{QPC1}$. The phenomenological binomial factor $(1 - R_\mathrm{QPC1})$ in Eq. (1) relates to charge fluctuation of non-interacting particles in the QPC. The temperature-dependent term emanates from the detailed balance principle [22].

## Time-Domain Braiding by Diluted Beam

We extend Eq. (1) to the two-QPC configuration. When a *diluted* beam is partitioned by QPC2, the spectral density $S_\mathrm{QPC2}$ of the excess autocorrelation of current fluctuations in QPC2 can be expressed as,

$$S_\mathrm{QPC2} = \mathcal{F}_\mathrm{dilute} \times 2e^*I_\mathrm{QPC1}R_\mathrm{QPC2}\big(1 - R_\mathrm{QPC2}\big)\left[\coth\left(\frac{e^*V_\mathrm{S}}{2k_\mathrm{B}T}\right) - \frac{2k_\mathrm{B}T}{e^*V_\mathrm{S}}\right], \qquad (2)$$

with $\mathcal{F}_\mathrm{dilute}$ being dependent on the diluting $R_\mathrm{QPC1}$ of the beam (Supplementary Note III), and $R_\mathrm{QPC2}$ is the reflection probability of QPC2. This expression has the same structure as Eq. (1), with the replacement of $I_\mathrm{S}$ with $I_\mathrm{QPC1}$ and $R_\mathrm{QPC1}$ with $R_\mathrm{QPC2}$. In the limit of large $V_\mathrm{S}$ and small $R_\mathrm{QPC2}$, it becomes $S_\mathrm{QPC2} = \mathcal{F}_\mathrm{dilute} \times 2e^*I_\mathrm{QPC2}$ with the current $I_\mathrm{QPC2} = I_\mathrm{QPC1}R_\mathrm{QPC2}$. It is note that for free fermions $\mathcal{F}_\mathrm{dilute} = 1$.

The Fano factor $\mathcal{F}_\mathrm{dilute}$ distinguishes between different partitioning processes. We consider the limits of large $V_\mathrm{S}$ and small $R_\mathrm{QPC2}$, where $I_\mathrm{QPC2} = e^*(W_{2\to3} - W_{3\to2})$, with spectral density $S_\mathrm{QPC2} = 2e^{*2}(W_{2\to3} + W_{3\to2})$, and $\mathcal{F}_\mathrm{dilute} = (W_{2\to3} + W_{3\to2})/(W_{2\to3} - W_{3\to2})$. Among possible partitioning processes, we first consider the trivial partitioning where an anyon in the dilute beam directly tunnels at QPC2 from Edge2 to Edge3 (Fig. 2(a)). This ubiquitous partitioning manifests particle antibunching [4-6], regardless of whether the particle is a boson, a fermion or an anyon. Here $\mathcal{F}_\mathrm{dilute} = 1$ as the rate $W_{2\to3}$ exponentially dominates $W_{3\to2}$ at high enough voltage ($e^*V_\mathrm{S} \gg k_\mathrm{B}T$), in a similar fashion to the partitioning of a full beam.

However, the trivial partitioning process of a highly diluted anyonic beam with a high source voltage $V_S$ leads to only a subdominant contribution to the observables. Instead, a more dominant process, which involves anyon braiding, takes place [7, 9]. In this process, which we call time-domain braiding, the anyon that tunnels between Edge2 and Edge3 (for example, from Edge2 to Edge3) at time $t_1$, leaving a hole behind (on Edge2). This anyon tunnels back at time $t_2$ and is 'pair-annihilated' with the hole as long as $t_2 - t_1 \lesssim \hbar/k_\mathrm{B}T$, where $\hbar$ is the reduced Planck constant. These probabilistic events of the particle-hole excitation and recombination form a loop in the time-domain. The time-domain loop of the thermal anyon in QPC2 braids with the anyons in the diluted beam that arrive at QPC2 during the time interval $t_2 - t_1$ (Fig. 2(b)), thus gaining a braiding phase (see below). The time-domain braiding dominates over the trivial partitioning as, according to the CLL theory, anyon tunnelling at a QPC becomes suppressed at higher energy. Within QPC2, anyon tunneling for a thermal particle-hole pair excitation (with energy approximately $k_\mathrm{B}T$) happens much more frequently than the tunneling of an arriving diluted anyons (with energy approximately $e^*V_\mathrm{S} \gg k_\mathrm{B}T$, and required for the trivial partition).

Being fundamental in our experiment, we stress the time-domain braiding process again. The thermal particle-hole excitation happens at QPC2 between Edge2 and Edge3 either before (at $t_1$) or after (at $t_2$) the arrival of the diluted anyons at QPC2. These two subprocesses differ by an exchange phase, as the spatial order of the anyons (the thermal particle-hole and the arriving dilute anyons) on Edge2 differs between the sub-processes (Supplementary Fig. S9). The interference between the subprocesses forms the time-domain loop of the thermal anyons that braids the diluted anyons. This braiding process leads to a modified Fano factor $\mathcal{F}_{\text{dilute}}$ (Methods and Supplementary Note III),

$$\mathcal{F}_{\text{dilute}} = -\cot\pi\delta\cot\left(\left(\frac{\pi}{2}-\theta\right)(2\delta-1)\right) \approx 3.27, \tag{3}$$

when $R_{\text{QPC1}} \ll 1$. Here, $\delta$ is the scaling dimension of anyon tunnelling at QPC2, and $2\theta$ ($\neq 0, 2\pi$) is the braiding angle. The value $\mathcal{F}_{\text{dilute}}$=3.27 is obtained with the ideal $\nu = 1/3$ state, with the corresponding $\delta = 1/3$ and $\theta = \pi/3$.

As measuring the excess autocorrelation of a highly diluted beam is challenging, we developed a phenomenological theory for a moderately diluted beam. Going beyond the CLL theory, the critical step is the identification of the average braiding phase in the time-domain braiding process

$$\langle e^{2ik\theta}\rangle_{\text{binomial}} = \sum_{k=0}^{n} P_k\, e^{2ik\theta} = \left(1 - R_{\text{QPC1}} + R_{\text{QPC1}}e^{2i\theta}\right)^n, \tag{4}$$

where $k$ denotes the number of anyons in the dilute beam which arrive at QPC2 in the time interval $t_2 - t_1$. The phase term $e^{2ik\theta}$ corresponds to the braiding phase of a thermally excited anyon with each of the arriving anyons. The probability $P_k$ of the $k$ anyon event is naturally assumed to follow the binomial distribution $P_k = \frac{n!}{k!\,(n-k)!}(R_{\text{QPC1}})^k(1 - R_{\text{QPC1}})^{n-k}$, that is, the probability for $k$ anyons being reflected by QPC1 with reflection probability $R_{\text{QPC1}}$. The maximum value of $k$ is $n = I_{\text{S}}(t_2 - t_1)/e^*$. The average braiding phase is implemented in the calculation of $\mathcal{F}_{\text{dilute}}$ using the ideal CLL parameters (as above) and integrating over the time difference $t_2 - t_1$. As the beam is less diluted (that is, fuller), the trivial partitioning process is also considered in the above expression, although its contribution is small (Methods and Supplementary Note III). It is note that the average braiding phase is $\langle e^{2ik\theta}\rangle_{\text{binomial}} = 1$ for fermions ($\theta = \pi$) and for bosons ($\theta = 0$).

## Experimental Results

We measured the excess spectral density $S_{\text{B}}$ of the excess autocorrelation for two partitioning cases: injection of a *full* beam and injection of a *dilute* beam. We first performed these measurements in the integer regime (the outer edge mode of filling factor $\nu = 3$). The Fano factors in both cases agree with trivial partitioning $\mathcal{F}_{\text{dilute}} = 1$, with the expected electronic charge $e^* = e$ (Supplementary Note II). Similar measurements were performed at filling $\nu = 1/3$. Injecting a *full* beam led to $S_{\text{B}}$ agreeing with Eq. (1) with charge $e^* \approx e/3$ (Supplementary Note II). Injecting a *dilute* beam, with $R_{\text{QPC1}}, R_{\text{QPC2}} \approx 0.1 \ll 1$, the experimental values of $\mathcal{F}_{\text{dilute}}$ were found close to $\mathcal{F}_{\text{dilute}} \approx 3.27$ (Eqs. (3 and 4) and Fig. 3); ruling out the trivial process ($\mathcal{F}_{\text{dilute}} = 1$) and substantiating the time-domain braiding process. Here we utilized that $S_{\text{B}}$ coincides with $S_{\text{QPC2}}$ at large voltages (Supplementary Note IV).

In Fig. 4, the spectral density $S_{\text{B}}$ of the autocorrelation was measured with varying dilutions, $R_{\text{QPC1}}$, and different partitioning, $R_{\text{QPC2}}$. With less dilution ('fuller' beam), the time-domain braiding process gives rise to smaller $\mathcal{F}_{\text{dilute}}$ and the trivial partitioning contribution to $\mathcal{F}_{\text{dilute}}$ is higher, albeit still small. Notice the excellent agreement between the experimental data and the phenomenological theory over a wide range of $V_{\text{S}}/T$, $R_{\text{QPC1}}$ and $R_{\text{QPC2}}$, without fitting parameters. The deviation of the data from the theory at large $V_{\text{S}}$, compounded with less dilution (larger $R_{\text{QPC1}}$), is probably due to the variation of the QPC reflection with the source voltage $V_{\text{S}}$ (not taken into account in the theory).

Time-domain braiding requires coherence between the two subprocesses [7, 9]. The agreement between the experimental data and the theory with $2\theta = 2\pi/3$ and $\delta = 1/3$ in Figs. 3 and 4 implies that the inter-QPC distance of 2 μm is indeed shorter than the phase coherence length, and edge reconstruction [24] does not take place. To test this assumption, we fabricated a similar two-QPC geometry with an inter-QPC distance of 20 μm. In this case, the measured $S_{\text{B}}$ showed a clear deviation from $\mathcal{F}_{\text{dilute}} \approx 3.27$, following the trivial formalism of Eq. (1) for non-interacting particles, with $R_{\text{QPC1}} \rightarrow R_{\text{QPC1}} R_{\text{QPC2}}$ [25] (Fig. 5).

We extended our study to the fraction $\nu = 2/5$ (Supplementary Note V). Partitioning dilute anyons with $e^* = e/3$ (the outermost edge mode) at QPC2, we find a Fano factor close to $\mathcal{F}_{\text{dilute}} \approx 3.27$, which supports the time-domain braiding with $2\theta = 2\pi/3$ and $\delta = 1/3$ as in $\nu = 1/3$. However, partitioning with QPC2 the inner edge mode (conductance $e^2/15h$), carrying charge $e^* = e/5$, we found $\mathcal{F}_{\text{dilute}} \approx 1$, which is in our measurement's uncertainty ($R_{\text{QPC1}} = 0.088$, $R_{\text{QPC2}} = 0.186$). The result is close to the Fano factor corresponding to the trivial partition process (see above).

## Promise of Time-Domain Braiding

It might be worthwhile to compare our two-QPC configuration with a recent work based on a three-QPC setup [15]. In the latter work, the measured cross-correlation (of partitioned diluted 1/3-filling beams) agreed with quantum calculations [9, 19], and was attributed to 'anyon-bunching by collision' following a classical lattice model [19]. The collision is a different process from the time-domain braiding, providing only a subdominant contribution to the cross-correlation (similarly to trivial partitioning) [9]. In the collision process, two diluted anyons, injected from two side QPCs, simultaneously arrive at the central QPC and the presence of one anyon alters the tunneling of the other one (at the central QPC) owing to anyonic bunching. Consequently, we tested our theory by performing a three-QPC experiment and found the results to agree well with our phenomenological approach (at a relatively large $R_{\text{QPC1}}$), supporting the underlying physics of the time-domain anyon braiding (Supplementary Note VI). Therefore, we believe that the previous three-QPC experimental results [15] should be regarded as time-domain braiding rather than anyons bunching. We note that two recent experiments also support the time-domain braiding process [26, 27].

Here we demonstrate a relatively simple experimental configuration that identifies the statistical phase of abelian anyons in the FQH states. Our findings are also substantial considering the long-time disagreements between experiments (conductance and shot noise) and the chiral Luttinger theory [28]. For example, the theoretical voltage dependence of reflection probability in a QPC, $R_{\text{QPC}} \propto V^{2\delta-2}$, has not been confirmed experimentally (Supplementary Note II). As such, it is worth examining the robustness of our Fano factor, $\mathcal{F}_{\text{dilute}}$, with respect to a variation in the scaling dimension $\delta$. We find that $\mathcal{F}_{\text{dilute}}$ is expected to vary only by 10% throughout the range $1/3 < \delta < 2/3$ (Supplementary Note III).

Although it is natural to expect that a highly diluted particle beam, such as photons or electrons [4-6], exhibits single-particle scattering at a barrier, our work shows an exception to this expectation: impinging highly diluted fractional QPs undergo multi-particle scattering at a QPC constriction, as they are topologically linked (braided) with the time-domain trajectory of thermally excited anyons within the constriction. This feat is accomplished by a relatively simple two-QPC configuration – allowing a straightforward identification of the braiding phase in a considerably simpler method than interference experiments. Moreover, our work suggests a promising route towards observing the topological order of non-abelian anyons, such as in the 5/2 filling in the FQH regime [9].

## METHODS

### Theory of the Fano factor

In the CLL theory and the equivalent conformal field theory [9], the time-domain braiding process is described by a non-equilibrium correlator $C_{\mathrm{neq}}(t_1, t_2)$ of the anyon tunnelling operator at QPC2 in the presence of a dilute anyon beam impinging at QPC2. It is expressed as $C_{\mathrm{neq}}(t_1, t_2) = \langle e^{2ik\theta}\rangle_{\mathrm{Poissonian}} C_{\mathrm{eq}}(t_1, t_2)$ in the limit of a highly diluted beam, namely, $R_{\mathrm{QPC1}} \ll 1$, where $C_{\mathrm{eq}}(t_1, t_2)$ is the equilibrium correlator in the absence of the dilute beam. Here, $\langle e^{2ik\theta}\rangle_{\mathrm{Poissonian}} = \sum_{k=0}^{\infty} Q_k\, e^{2ik\theta}$ is the average of the braiding phase $e^{2ik\theta}$, which accumulates when the time-domain loop of thermally excited anyons braids with $k$ anyons of the dilute beam arriving at QPC2 in the time interval $t_2 - t_1$. The probability $Q_k$ represents $k$ random anyon injections from Edge1 to Edge2 at QPC1 (Figs. 1 and 2) over the time interval $t_2 - t_1$. For a highly diluted beam the Poisson distribution is $Q_k = (m^k/k!)e^{-m}$, where $m = I_{\mathrm{QPC1}}(t_2 - t_1)/e^*$.

It is naturally expected that in a less dilute ('fuller') beam (with a relatively large $R_{\mathrm{QPC1}}$, yet small enough for anyon tunnelling), the distribution of anyons in the beam follows a binomial distribution rather than the Poissonian distribution. Hence, to describe the cases of less dilute beams, we replace the multiplicative factor $\langle e^{2ik\theta}\rangle_{\mathrm{Poissonian}}$ by the average braiding phase $\langle e^{2ik\theta}\rangle_{\mathrm{binomial}}$, with the latter averaged over the binomial distribution in Eq. (4). Then the correlator is,

$$C_{\mathrm{neq}}(t_1, t_2) = \left(1 - R_{\mathrm{QPC1}} + R_{\mathrm{QPC1}} e^{2i\theta \mathrm{sign}(t_2 - t_1)}\right)^{\frac{I_S}{e^*}|t_1 - t_2|} C_{\mathrm{eq}}(t_1, t_2). \quad (5)$$

In the dilute limit of $R_{\mathrm{QPC1}} \ll 1$, the multiplicative factor $(1 - R_{\mathrm{QPC1}} + R_{\mathrm{QPC1}} e^{\pm 2i\theta})^{\frac{I_S}{e^*}|t_1 - t_2|}$ is reduced to the factor $e^{-(1 - e^{\pm 2i\theta})\frac{I_{\mathrm{QPC1}}}{e^*}|t_1 - t_2|}$ found in a previous work [9]. Employing $C_{\mathrm{neq}}(t_1, t_2)$ with an integral over $t_2 - t_1$, it is straightforward to compute the rates of anyon tunnelling (back and forth) at QPC2 in the time-domain braiding process. At zero temperature and $R_{\mathrm{QPC2}} \ll 1$, we get

$$W_{2\to 3}^{\mathrm{braid}} \propto \mathrm{Re}\left[e^{i\pi\delta}\left(-\log\left(1 + R_{\mathrm{QPC1}}\left(e^{-i2\theta} - 1\right)\right)\right)^{2\delta - 1}\right],$$
$$W_{3\to 2}^{\mathrm{braid}} \propto \mathrm{Re}\left[e^{i\pi\delta}\left(-\log\left(1 + R_{\mathrm{QPC1}}\left(e^{i2\theta} - 1\right)\right)\right)^{2\delta - 1}\right], \quad (6)$$

with the full expressions given in Supplementary Note III. In contrast to the trivial process where $W_{3\to2}$ is exponentially suppressed in comparison with $W_{2\to3}$, both $W_{2\to3}^{\mathrm{braid}}$ and $W_{3\to2}^{\mathrm{braid}}$ are non-negligible in the time-domain braiding. The appearance of the combination ($e^{\pm i2\theta}-1$) in Eq. (6) implies that the rates $W_{2\to3}^{\mathrm{braid}}$ and $W_{3\to2}^{\mathrm{braid}}$ vanish, and thus do not contributing to the tunnelling currents and noise at QPC2 in the cases of fermions ($\theta=\pi$) or bosons ($\theta=0$). Hence the time-domain braiding does not exist with fermions or bosons, but only with anyons [29-32].

When the time-domain braiding process dominates over other processes, the Fano factor is written as

$$\mathcal{F}_{\mathrm{dilute}}=\frac{W_{2\to3}^{\mathrm{braid}}+W_{3\to2}^{\mathrm{braid}}}{W_{2\to3}^{\mathrm{braid}}-W_{3\to2}^{\mathrm{braid}}}=-\cot\pi\delta\,\frac{\mathrm{Re}\left[\left(-\log\left(1+R_{\mathrm{QPC1}}\left(e^{-i2\theta}-1\right)\right)\right)^{2\delta-1}\right]}{\mathrm{Im}\left[\left(-\log\left(1+R_{\mathrm{QPC1}}\left(e^{-i2\theta}-1\right)\right)\right)^{2\delta-1}\right]}. \quad (7)$$

In the dilute limit of $R_{\mathrm{QPC1}}\ll1$, we find $\mathcal{F}_{\mathrm{dilute}}\to-\cot\pi\delta\frac{\mathrm{Re}[(1-e^{-i2\theta})^{2\delta-1}]}{\mathrm{Im}[(1-e^{-i2\theta})^{2\delta-1}]}$ as in Eq. (3). That zero-temperature value of $\mathcal{F}_{\mathrm{dilute}}$ of the two-QPC set-up corresponds to the Fano factor of the cross-correlation of a three-QPC set-up predicted in Ref. [9, 19]. As the beam becomes less dilute ('fuller'), the trivial partitioning process contributes more to the rates of $W_{2\to3}^{\mathrm{triv}}$ and $W_{3\to2}^{\mathrm{triv}}$ (Supplementary Note III). Then the Fano factor $\mathcal{F}_{\mathrm{dilute}}$ is obtained according to all the rates accounted for all the processes, $W_{2\to3}=W_{2\to3}^{\mathrm{braid}}+W_{2\to3}^{\mathrm{triv}}$ and $W_{3\to2}=W_{3\to2}^{\mathrm{braid}}+W_{3\to2}^{\mathrm{triv}}$, with the experimentally measured $R_{\mathrm{QPC1}}$ as input of the calculation. We note that $W_{2\to3}^{\mathrm{triv}}$ and $W_{3\to2}^{\mathrm{triv}}$ are not negligible but much smaller than $W_{2\to3}^{\mathrm{braid}}$ and $W_{3\to2}^{\mathrm{braid}}$ for the values of $R_{\mathrm{QPC1}}$ studied in our experiments.

It should be noted that partitioning a non-diluted beam can provide an anyonic signature through a different process from our partitioning a strongly diluted beam [33].

## Obtaining $S_{\mathrm{QPC1}}$ in a two-QPC configuration

While performing the two-QPC measurements, the noise generated by QPC1 ($S_{\mathrm{QPC1}}$) is not directly accessible (owing to the locations of the amplifiers). However, current conservation can be used to relate $S_{\mathrm{QPC1}}$ to the correlations measured in the experiment. By current conservation in QPC2,

$$I_{\mathrm{QPC1}}=I_{\mathrm{QPC2}}^{\mathrm{A}}+I_{\mathrm{QPC2}}^{\mathrm{B}},$$

where $I_{\mathrm{QPC1}}$ is the dilute current generated by QPC1 and $I_{\mathrm{QPC2}}^{\mathrm{A/B}}$ is the output current of QPC2 that reaches amplifier A/B (Fig 1(a)). The same relation also holds for the averages

$$\langle I_{\mathrm{QPC1}}\rangle=\langle I_{\mathrm{QPC2}}^{\mathrm{A}}\rangle+\langle I_{\mathrm{QPC2}}^{\mathrm{B}}\rangle.$$

Subtracting these two equations and taking the square we arrive at a relation between the current correlations

$$S_{\mathrm{QPC1}}=S_{\mathrm{A}}+S_{\mathrm{B}}+2S_{\mathrm{AB}},$$

which allows us to obtain $S_{\mathrm{QPC1}}$ by summing the autocorrelations ($S_{\mathrm{A}}$ and $S_{\mathrm{B}}$) and the cross-correlation ($S_{\mathrm{AB}}$) measured in the experiment.

## Author's contribution

J.-Y.M.L. and H.-S.S. developed the theory and analyzed the data. C.H. and T.A. fabricated the structures, did all measurements and analyzed the data. N.S. and Y.O. added perspective on the theory. V.U. designed and grew the heterostructures by Molecular Beam Epitaxy. M.H. supervised the experiments.

## Acknowledgments

J.-Y.M.L. acknowledges support from Korea NRF, NRF-2019-Global Ph.D. fellowship. H.-S.S acknowledges support from Korea NRF, the SRC Center for Quantum Coherence in Condensed Matter (Grant No. 2016R1A5A1008184). N.S. acknowledges fruitful discussions with Yotam Shapira and Ady Stern, and. acknowledges the Clore Scholars Programme. Y.O. acknowledges the partially supported by grants from the ERC under the European Union's Horizon 2020 research and innovation programme (grant agreements LEGOTOP No. 788715 and HQMAT No. 817799), the DFG (CRC/Transregio 183, EI 519/7-1), the BSF and NSF (2018643), the ISF Quantum Science and Technology (2074/19). M.H. acknowledges the continuous support of the Sub-Micron Center staff, the support of the European Research Council under the European Community's Seventh Framework Program (FP7/2007-2013)/ERC under grant agreement number 713351. We thank Gwendal Fève, Frédéric Pierre, and Christophe Mora for their insightful discussions.

## References for the Methods

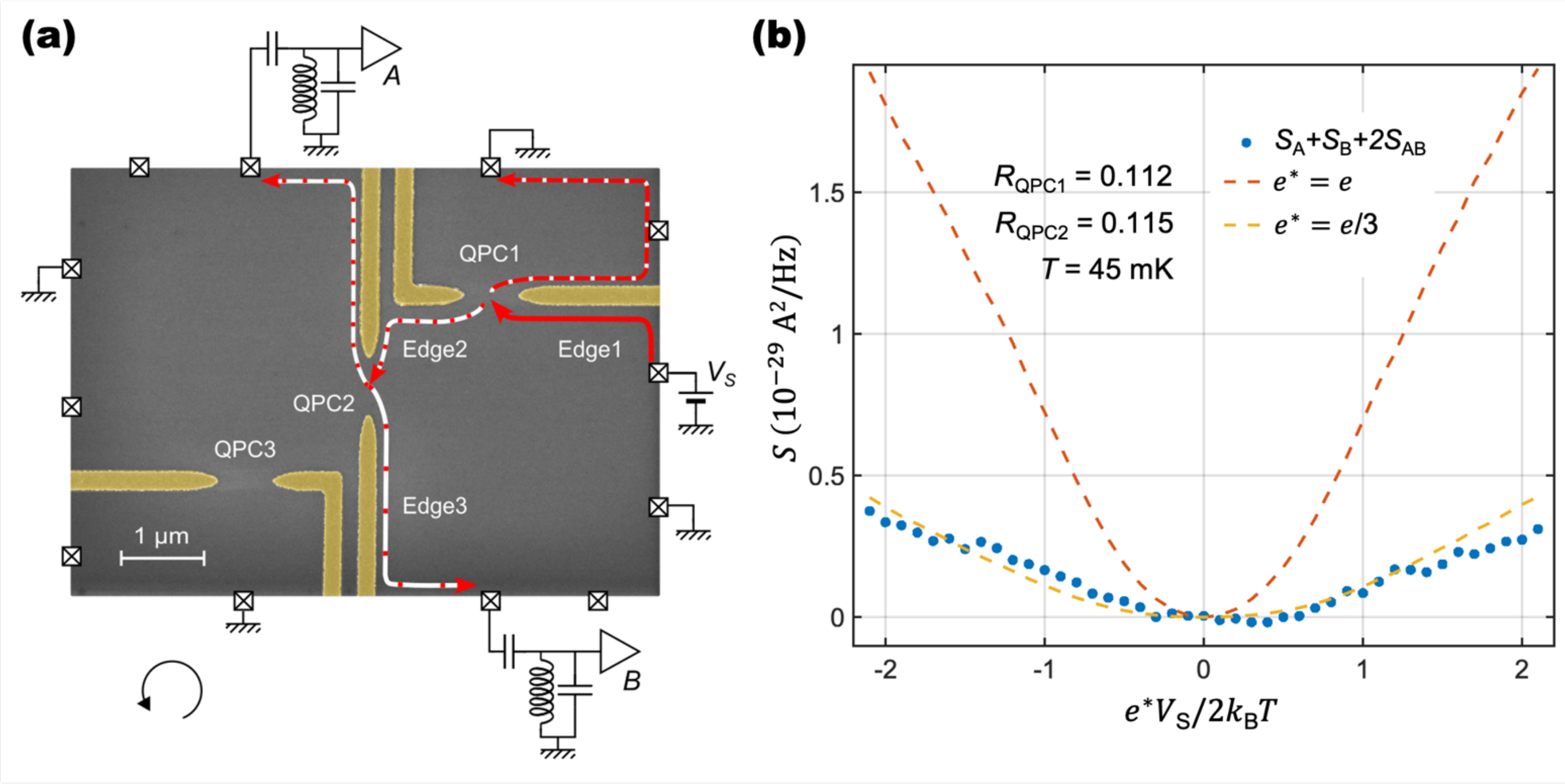

**Figure 1. Partitioning diluted anyons in a two-QPC geometry. (a)** The experimental set-up. False-colour SEM image with edge modes. The metallic gates of the QPC are coloured yellow. The ohmic contacts are more than 100 μm away from the core structure. The source current propagates along Edge1 and is diluted by QPC1 with $R_{QPC1}$. The diluted beam reaches QPC2 fabricated 2 μm away along Edge2. Partitioning takes place in QPC2 with back reflection along Edge3. The two amplifiers measure the excess autocorrelations and the cross-correlation. **(b)** The spectral density of the noise generated by QPC1, with charge $\boldsymbol{e^*} = \boldsymbol{e/3}$ (blue dots – data; yellow dashed line - expected). It is obtained by a summation of the autocorrelations and cross-correlation of the current fluctuation in QPC2 diluted by QPC1 (Methods). Using this method, the injected QP charge towards QPC2 was found to be $\boldsymbol{e/3}$. The experimental parameters are shown on the top left (detail in Supplementary Note II). The expected shot noise for a charge $\boldsymbol{e^*} = \boldsymbol{e}$ is shown for comparison (red dashed line).

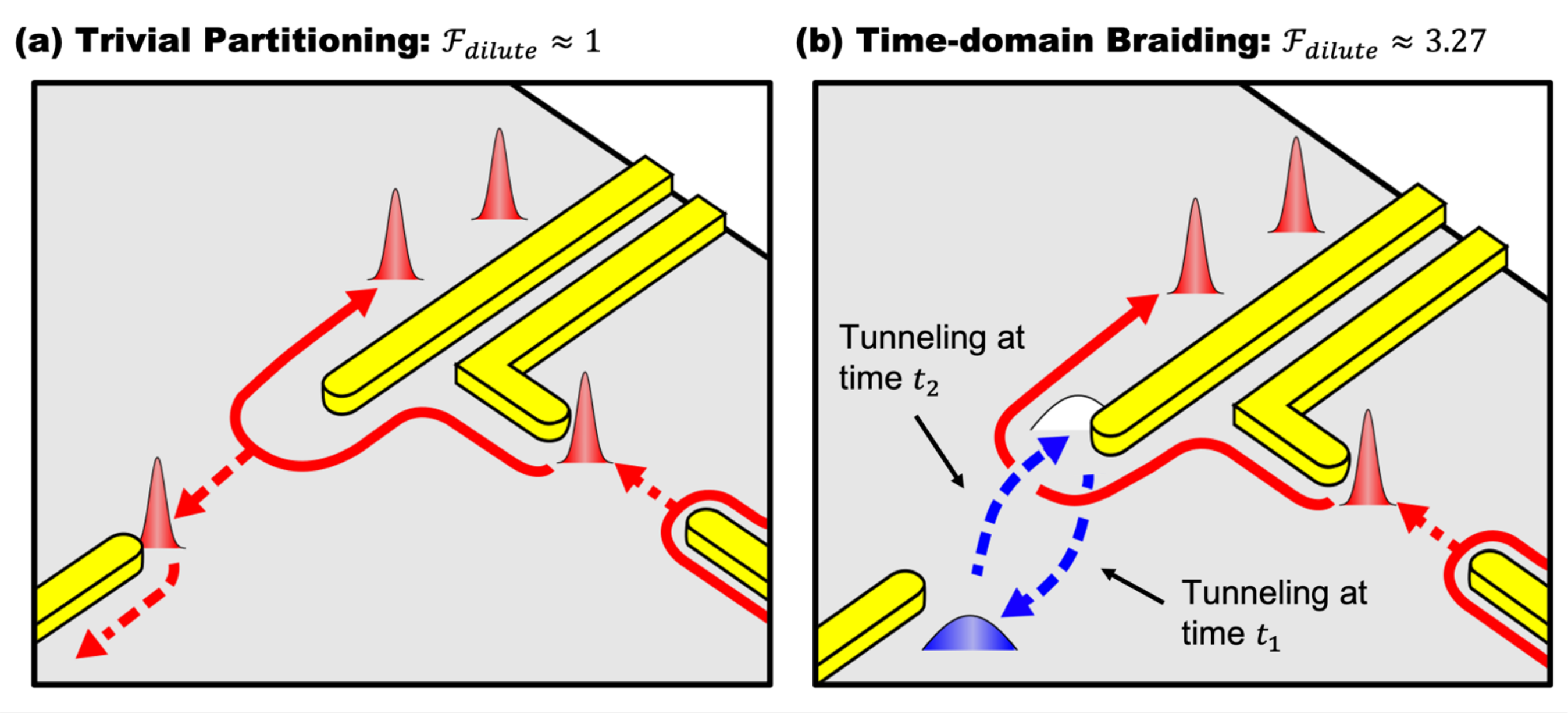

**Figure 2. Trivial and braiding partitioning processes in QPC2. (a)** Trivial partitioning: QPC1 dilutes the incoming beam by reflection $R_{QPC1}$ (red wavepackets), which is partitioned further in QPC2 by $\boldsymbol{R_{QPC2}}$. Shot noise is proportional to $\boldsymbol{R_{QPC1} R_{QPC2}}$. **(b)** Time-domain braiding: QPC1 dilutes the incoming beam by reflection $\boldsymbol{R_{QPC1}}$ (red wavepackets). A thermally activated particle-like anyon, depicted by a blue wavepacket, (leaving a hole, a white wavepacket) tunnels within QPC2 (blue arrow from one edge mode to another) at time $\boldsymbol{t_1}$. The diluted anyon arrived (with probability $\boldsymbol{R_{QPC1}}$). The particle-anyon tunnels back at a later time $\boldsymbol{t_2}$ (blue dashed arrows), thus braiding the arriving diluted anyon during the interval time $\boldsymbol{t_2 - t_1}$.

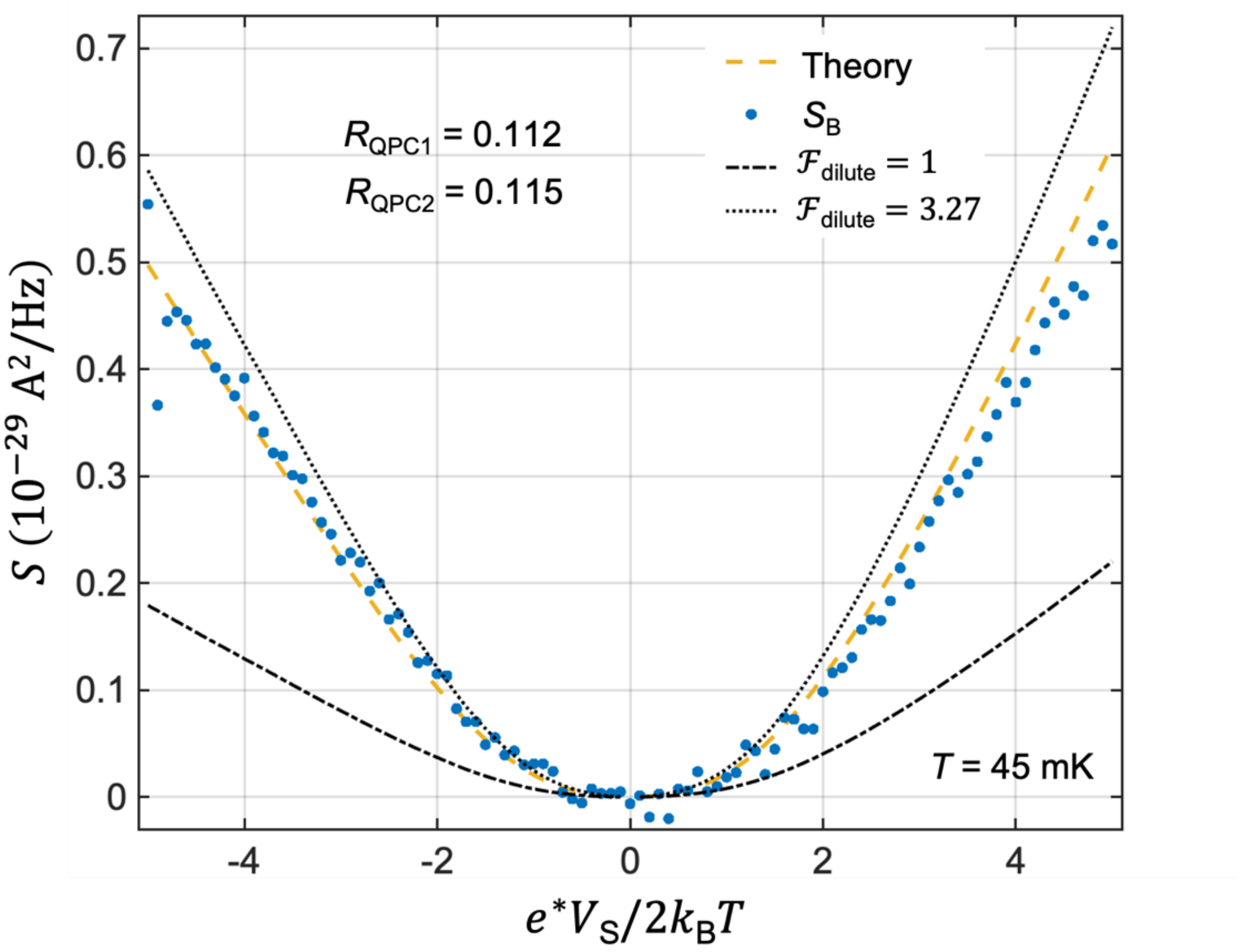

**Figure 3**. **Excess autocorrelation noise as measured at amplifier B (see Fig. 1a)**. A diluted beam of anyons is generated by reflection from QPC1 with probability $R_{\mathrm{QPC1}} = 0.112$. The dilute beam impinges on QPC2 with $R_{\mathrm{QPC2}} = 0.115$, creating excess autocorrelation (shot noise), shown by the blue dots. The yellow dashed line corresponds to the prediction of the phenomenological model given by Eq.(2), where $\mathcal{F}_{\mathrm{dilute}}$ is calculated based on the measured $R_{\mathrm{QPC1}}$ (Supplementary Note II and III). The black dotted line corresponds to the time-domain braiding process that dominates over the trivial process, with the Fano factor $\mathcal{F}_{\mathrm{dilute}} = 3.27$ (in the dilute limit $R_{\mathrm{QPC1}} \ll 1$). The black dashed line corresponds to $\mathcal{F}_{\mathrm{dilute}} = 1$, namely, the predicted noise of trivial partitioning in QPC2.

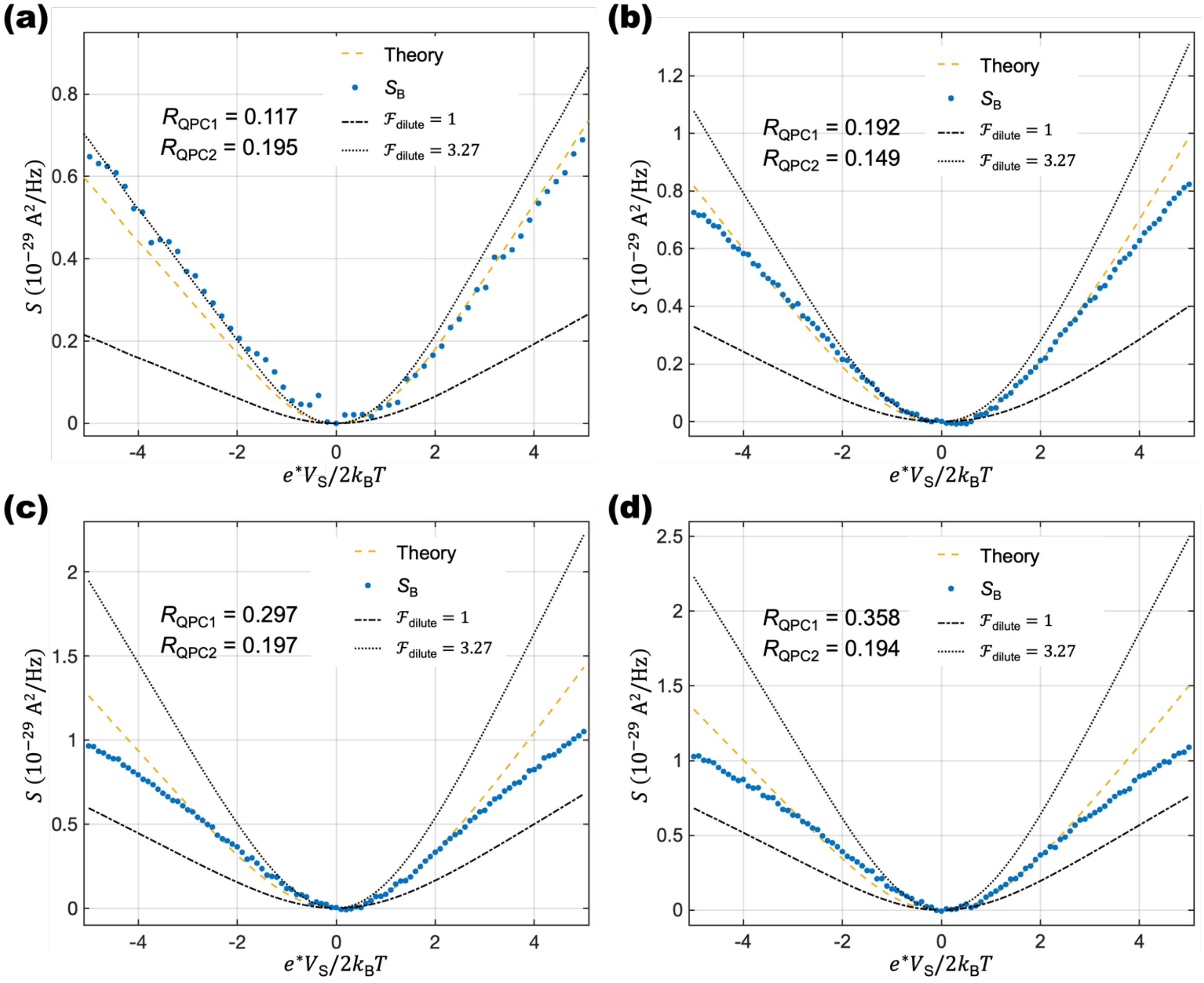

**Figure 4. The dependence of the autocorrelation (amplifier B, Fig. 1a) on beam dilution ($R_{QPC1}$) and on $R_{QPC2}$. (a)-(d)** From more to less dilution via $R_{QPC1}$, with **(a)** $R_{QPC1} = 0.117$, **(b)** $R_{QPC1} = 0.192$, **(c)** $R_{QPC1} = 0.297$, **(d)** $R_{QPC1} = 0.358$; excess autocorrelation (shot noise, blue dots) in the two-QPC configuration for different values of beam dilution. The yellow dashed lines are the theoretical predictions according to the phenomenological theory of Eq. (2). The black dashed lines are for the trivial process. The black dotted lines are for the dilute limit where the primary contribution to the noise results from the time-domain braiding process. The data are in a good agreement with the theory over a wide range of parameters. As predicted by Eq. (3), a higher dilution (smaller $R_{QPC1}$) minimizes the contribution of the trivial partitioning to the data, allowing the Fano factor of the autocorrelation to reach $\mathcal{F}_{dilute} = 3.27$. See also Supplementary Fig. S5.

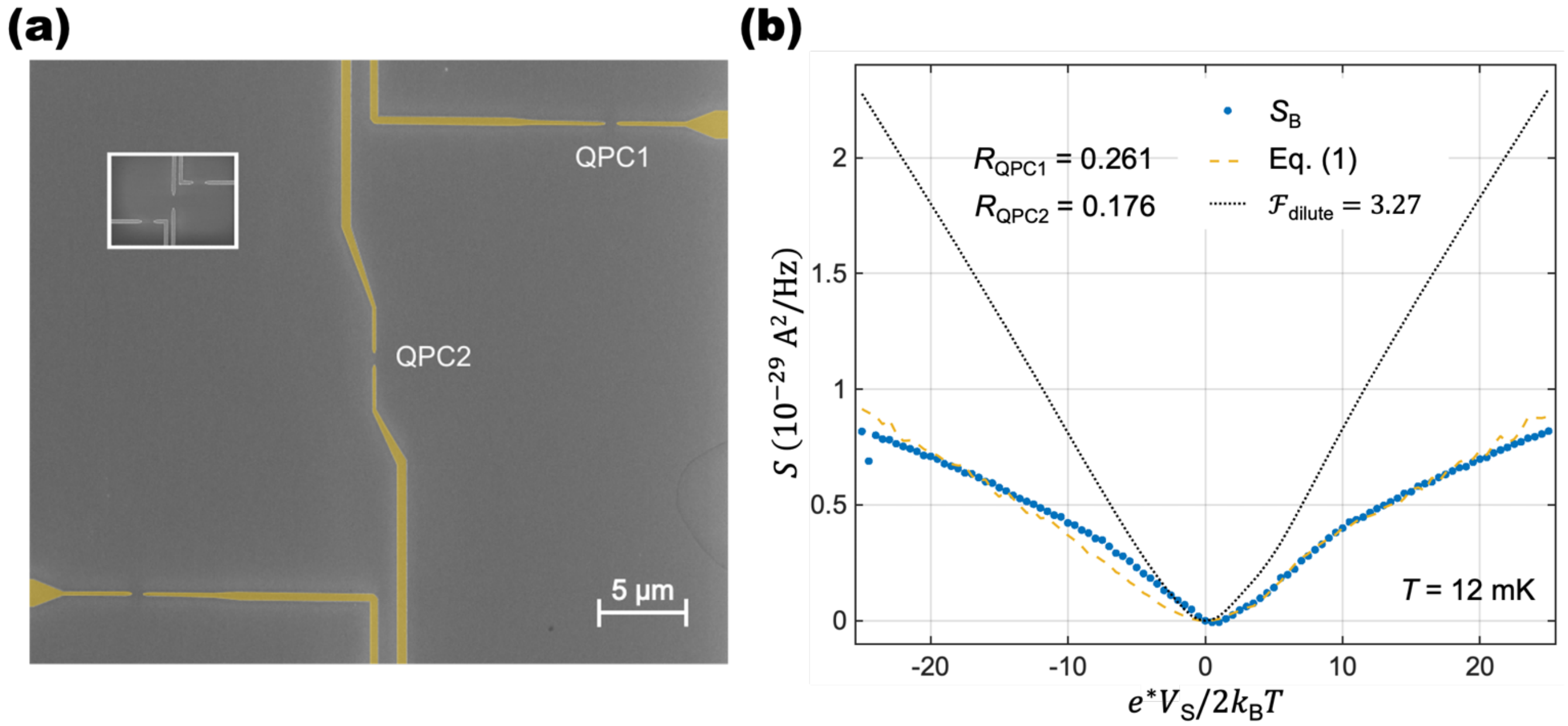

**Figure 5. Two-QPC configuration with an inter-QPC distance of 20μm. (a)** Scanning electron microscope image of the experimental set-up. The gates are marked in yellow. The 2-μm QPC separation structure is shown (for comparison) in the white-bordered inset. **(b)** The blue dots are the measured excess autocorrelation with dilution of $\boldsymbol{R_{QPC1}} = \mathbf{0.261}$ and $\boldsymbol{R_{QPC2}} = \mathbf{0.176}$. The measurement results agree with the trivial model (that is, integer filling factor) in Eq. (1) with $\boldsymbol{R_{QPC1}} \rightarrow \boldsymbol{R_{QPC1}R_{QPC2}}$, suggesting energy loss and dephasing due to the long propagation distance. The black dotted line is the ideal anyonic behavior with Fano Factor $\boldsymbol{\mathcal{F}_{dilute}} = \mathbf{3.27}$.

# Supplementary Note for
# Partitioning of Diluted Anyons Reveals Their Braiding Statistics

June-Young M. Lee[1*], Changki Hong[2*], Tomer Alkalay[2*], Noam Schiller[3], Vladimir Umansky[2],

Moty Heiblum[2], Yuval Oreg[3], H.-S. Sim[1]

[1]Department of Physics, Korea Advanced Institute of Science and Technology, Daejeon 34141, South Korea

[2]Braun Center for Submicron Research, Department of Condensed Matter Physics, Weizmann Institute of Science, Rehovot 7610001, Israel

[3]Department of Condensed Matter Physics, Weizmann Institute of Science, Rehovot 7610001, Israel

[*]Equal Contribution

Correspondent Authors e-mail:
H.-S. Sim……hs_sim@kaist.ac.kr
Moty Heiblum….moty.heiblum@weizmann.ac.il

## SI. EXPERIMENTAL SETUP DETAILS

We employ a high mobility GaAs-AlGaAs heterostructure that supports a two-dimensional electron gas (2DEG) 125 nm below the surface. The 2DEG has an electron density of $9.2 \times 10^{10}$ cm$^{-2}$ and low temperature (4.2kelvin) dark mobility of $3.9 \times 10^{6}$ cm$^2$V$^{-1}$s$^{-1}$. Our experimental setup is shown in Fig. S1. Three Quantum Point Contacts (QPCs) were patterned in close proximity and served as beam-splitters, and ohmic contacts were used as Sources and Drains. In the two-QPC setup, QPC1 dilutes the DC current $I_S$ that is injected from source contact S1. The diluted reflected part of the current is then partitioned by QPC2. In the three-QPC setup, another source contact S2 is biased and injects DC current to the sample. The current from S2 is diluted by QPC3, and then injected towards QPC2. In both setups, the auto-correlation (AC), shot noise of each output beam, as well as the cross-correlation (CC) between the two outputs, are measured at a frequency of 730kHz (set by two separated the LC circuits). Each signal was amplified by a home-made preamplifier cooled to 4.2 K, which was followed by a room temperature amplifier. The output of the amplification chain was fed into a home-made analog cross-correlator circuit, which can multiplies each signal with itself (AC), or with a second signal (CC). The output voltage from the analog cross-correlator was measured by a digital multimeter. In order to calibrate the auto-correlation and cross-correlation measurements, we measured the shot noise (AC) of a full beam at an integer filling factor (outer – most edge mode at $\nu = 3$). This was performed by fully pinching QPC1 while source contact is biased, such that only the QPC2 partitioned the beam. Comparing the auto-correlations and cross-correlations with Eq. (1) (in the text) allowed us to calibrate our system. In addition, we repeated this measurement at $\nu = 1/3$, and made sure that both auto-correlations and the cross correlations leads to the correct fractional charge based on Eq. (1).

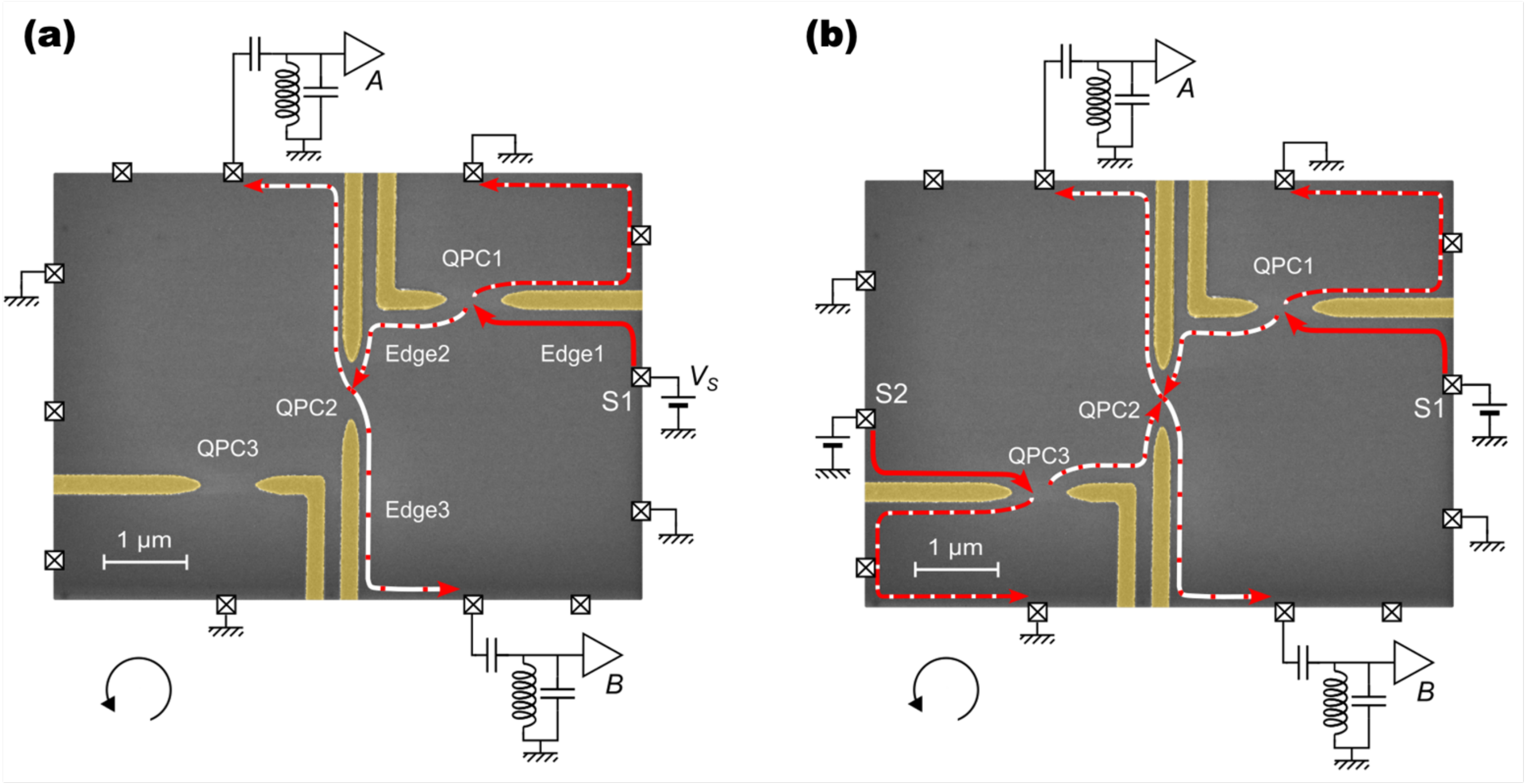

**FIG. S1**: **'Two-QPC' and 'three-QPC' configurations. (a)** A full beam is injected from source contact S1 and propagates along Edge1. The current is diluted by QPC1, the reflected part continues along Edge 2 and is partitioned by QPC2. **(b)** A second source (S2) is used to inject a full beam to QPC3, which is tuned such that it has the same tunneling probability as QPC1. The reflected current from QPC3 reaches the second input of QPC2. In both cases, the auto-correlation noise in each output beam is measured together with the cross-correlation between them.

## SII. SUPPLEMENTARY DATA

### A. Noise of a full beam impinging on QPC2

Here we compare the situation in which a dilute beam is injected to QPC2 (Fig. 3 in the main text) to that of a full beam injection. For this purpose, we utilized a second source contact [S2 contact in Fig. S1] to inject a full beam to QPC2, while QPC3 was fully pinched. QPC1 and QPC2 were held fixed at the same reflection used in the measurement of the dilute beam noise. As shown in Fig. S2, the noise follows Eq.(1) of the main text with charge $e^* \simeq e/3$. This measurement emphasizes the remarkable difference between a full beam injection to a QPC and the dilute injection. In the former as Fig. S2, the Fano factor is sensitive only to the partitioned charge dominated by the trivial partitioning. In the latter as Fig. 3 in the main paper, the time-domain braiding takes over and the Fano factor becomes dependent on the braiding phase of anyons.

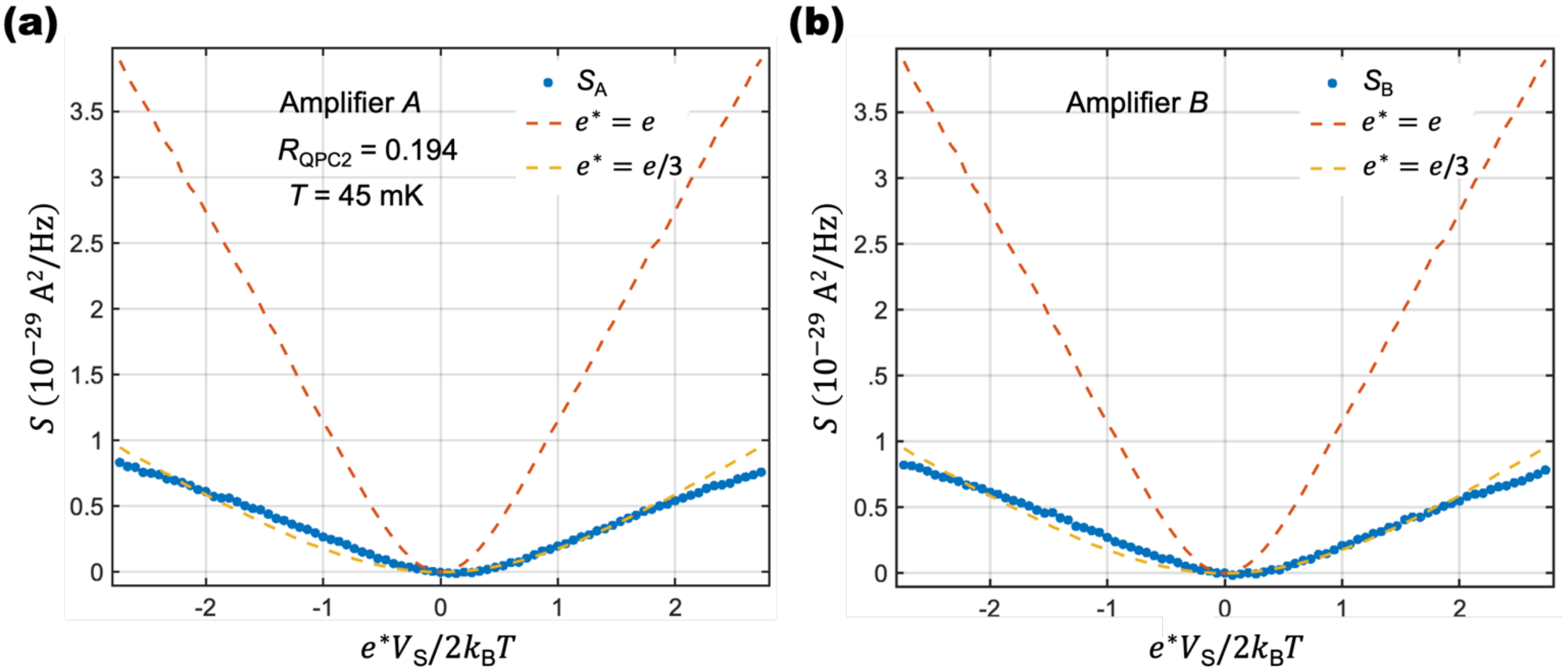

**FIG. S2: Noise of a full beam partitioned at QPC2.** A full beam is injected to QPC2 by biasing S2 and fully pinching QPC3. QPC1 and QPC2 were held at the same condition used to measure the noise of a dilute beam ($R_{QPC2}$=0.194). The AC noise at the two amplifiers (blue dots) is plotted together with the prediction of Eq. (1). The yellow dashed line is the expected noise with charge $e^* = e/3$ , while the red dashed line shows the expected noise for $e^* = e$ for comparison. The data is in very good agreement with Eq. (1), demonstrating that the noise of a full beam is only sensitive to the charge of the partitioned particles. This should be contrasted with the noise of a dilute beam, shown in the main paper, where the Fano factor becomes a probe of the statistical phase due to contributions from the time-domain braiding process.

### B. Two-QPC experiment in the IQH of filling factor 3

We performed the two-QPC experiment in the integer quantum Hall (IQH) regime, using the outer edge mode of filling factor 3. The condition is simpler (more ideal) to calculate because the DC bias dependency of the reflection probability is flat compared to other edge modes. In this regime, due to the trivial braiding phase of fermions, only the trivial partition process is expected to contribute to the noise, leading to $\mathcal{F}_{\mathrm{dilute}} = 1$. In Fig. S3, we compare the measured auto-correlation noise at amplifier A and B with Eq.(1) (dashed lines in Fig. S3), with the electronic charge $e^* = e$ and $R_{\mathrm{QPC1}}$ replaced by the total probability to reach the amplifier such that for amplifier A, namely, $R_{\mathrm{QPC1}}$ is replaced by $R_{\mathrm{QPC1}}(1 - R_{\mathrm{QPC2}})$, while for amplifier B, $R_{\mathrm{QPC1}}$ is replaced by $R_{\mathrm{QPC1}} R_{\mathrm{QPC2}}$. The theoretically expected values and the measurement results are in very good agreement, supporting that only the trivial partition process happens in the IQH.

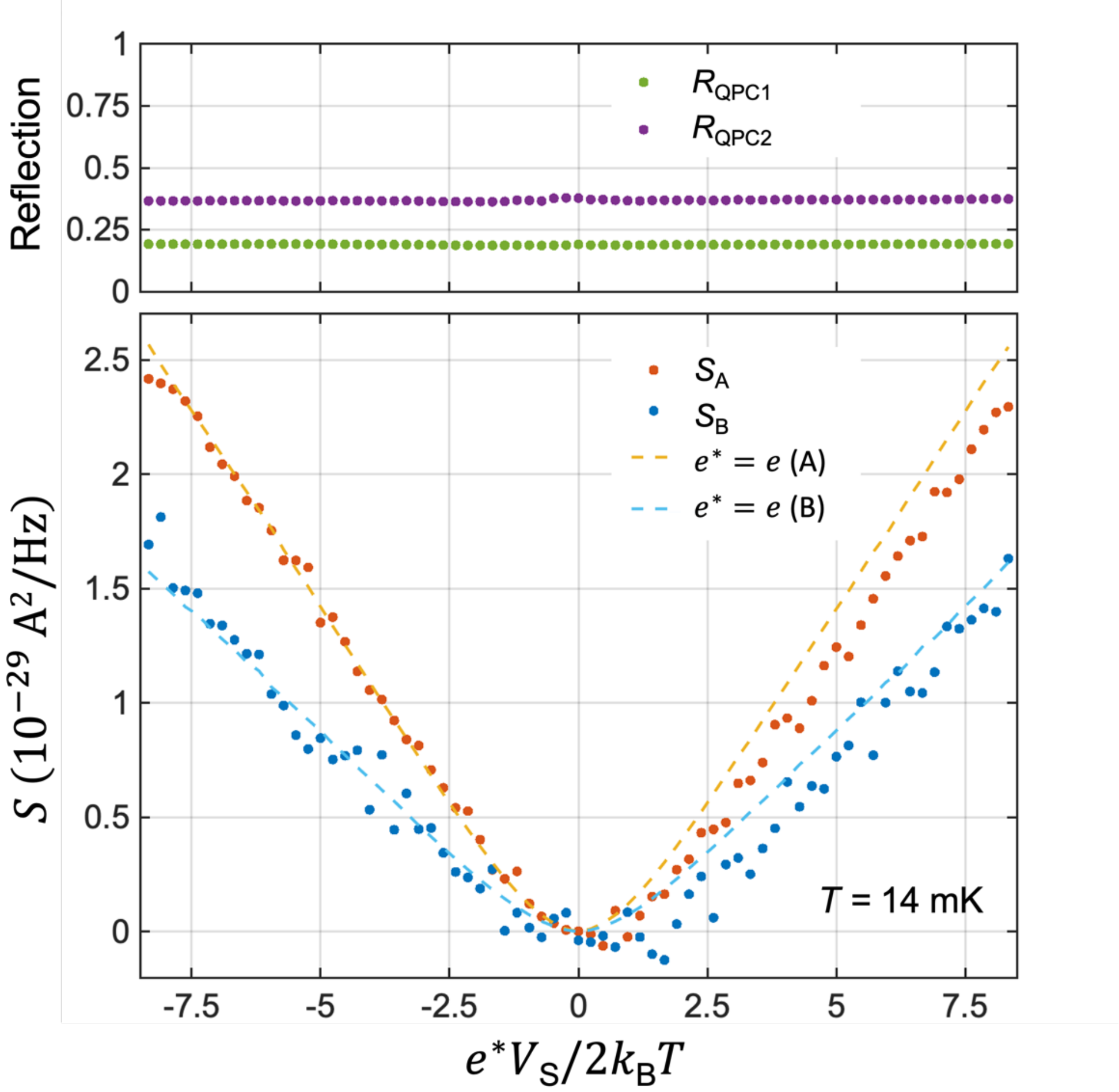

**FIG. S3: Excess auto-correlation noise in the two-QPC geometry at an integer filling factor ($\nu = 3$).** The upper panel shows the reflection probability of QPC1 and QPC2. In the lower panel, the red and blue dots are the measured the AC noises at the two amplifiers. The red dashed line is the noise expected by Eq. (1), with the electronic charge $e^* = e$ and $R_{\mathrm{QPC1}}$ is replaced by the combined probability to reach amplifier A, which is $R_{\mathrm{QPC1}}(1 - R_{\mathrm{QPC2}})$. Similarly, the blue dashed line is the expected noise at amplifier B with $e^* = e$ and $R_{\mathrm{QPC1}} \rightarrow R_{\mathrm{QPC1}} R_{QPC2}$. The agreement with the expected noise indicates that there is no additional contribution to the noise apart from the contribution of the trivial partitioning.

## C. Bias dependence of the reflection

In the main text, each noise measurement is shown along with the value of the reflection probability of each of the relevant QPCs averaged over the bias range. Here, we show the full bias dependence of the reflection probabilities which was used in generating the theoretical curves shown in the main text. Each panel of Fig. S4 shows the measured reflection probability for QPC1 ($R_{\mathrm{QPC1}}$) and QPC2 ($R_{\mathrm{QPC2}}$) and corresponds to one of the noise measurement presented in the main text. The corresponding noise measurement in the main text is written in the inset of each of the sub-figures. Figures S4 (a) to (e) measured at 45 mK and Figure S4 (f) measured at 12 mK.

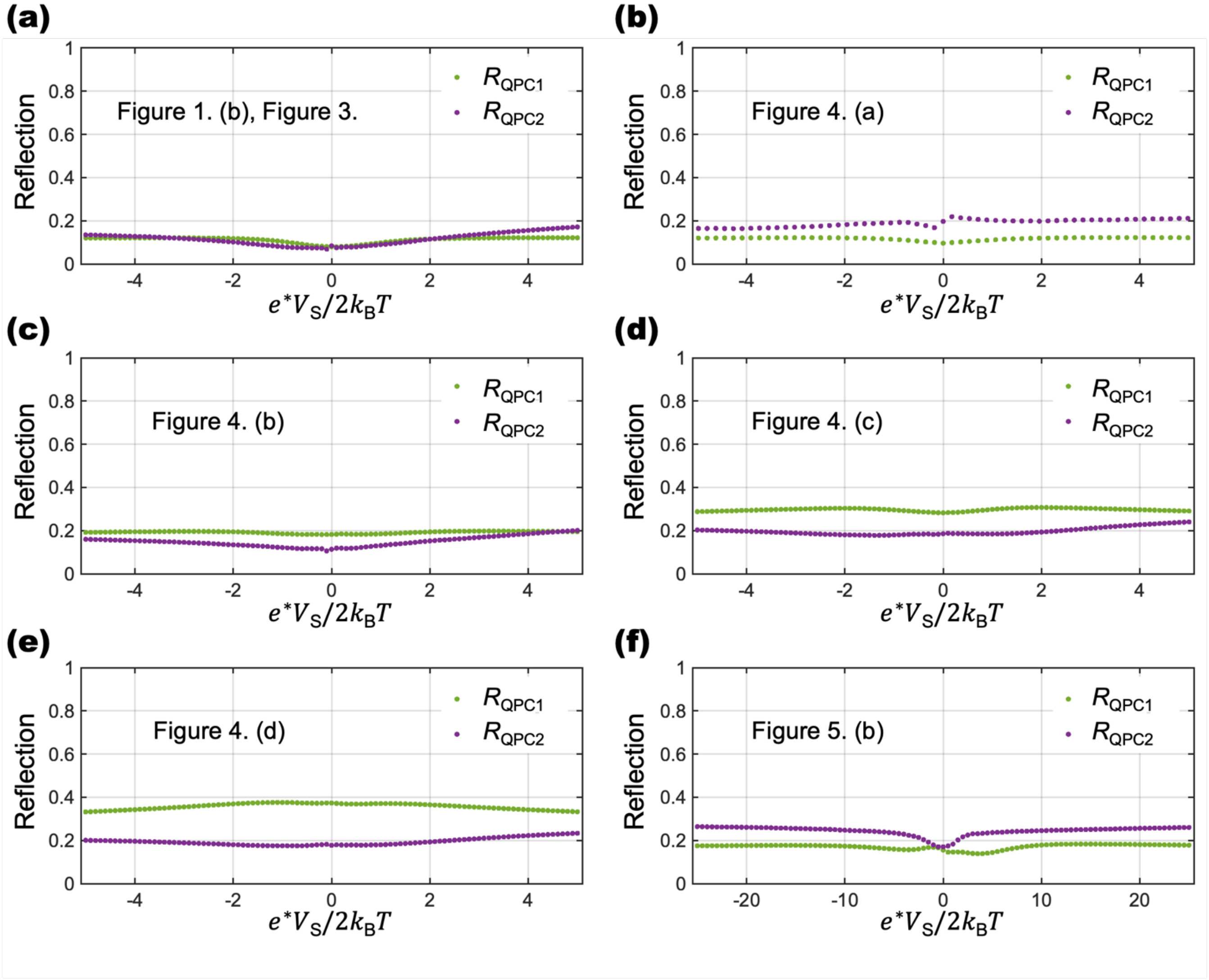

**FIG. S4: Bias dependence of the reflection probability.** Each of the panels **(a)**-**(f)** shows the full bias dependence of reflection probability measurement results for $R_{\mathrm{QPC1}}$ (green dots) and $R_{\mathrm{QPC2}}$ (purple dots). Each panel corresponds to noise measurement results in the main text: **(a)** corresponds to Fig. 1(b) and Fig. 3, **(b)** to Fig. 4(a), **(c)** to Fig. 4(b), **(d)** to Fig. 4(c), **(e)** to Fig. 4(d), and **(f)** to Fig. 5(b).

### D. Fitting of the exchange phase $\theta$

In Figs. 3 & 4 of the main text, the data is shown alongside the prediction of the phenomenological model for the ideal $\nu = 1/3$ case, with $\delta = 1/3$ and $\theta = \pi/3$. In this sub-section, we show the results of fitting the data to the theoretical prediction of the phenomenological model, with $\theta$ as a fitting parameter and $\delta = 1/3$. We used Eq. (2) together with the $\mathcal{F}_{\mathrm{dilute}}$ of the Eq. (7) in the main text, which means that we omit the contribution of the trivial partition process and employ the zero-temperature value of $\mathcal{F}_{\mathrm{dilute}}$ for simplicity [see Supplementary Note SIII for the trivial process' contribution and the temperature dependence]. In Fig. S5(a), we plotted the best fitted $\theta$ together with 95% confidence intervals as a function of the averaged beam dilution $R_{\mathrm{QPC1}}$. The uncertainty of each point results from a statistical uncertainty involved in measuring noise and a systematic uncertainty coming from the calibration process.

For the most dilute case [corresponding to Fig. 3] we find $\theta = 0.982 \pm 0.074$, and the corresponding fitting curve is shown in Fig. S5(b). The fitted value of $\theta$ deviates more from the ideal value of $\pi/3$ for the less dilute beams. It is partially because we did not include the trivial partition process in the fitting [the contribution from the trivial partition process becomes larger for a less dilute beam; see Eq. (S11)]. Also, the distortion of the QPC potential at the large voltage would be more severe for a less diluted beam.

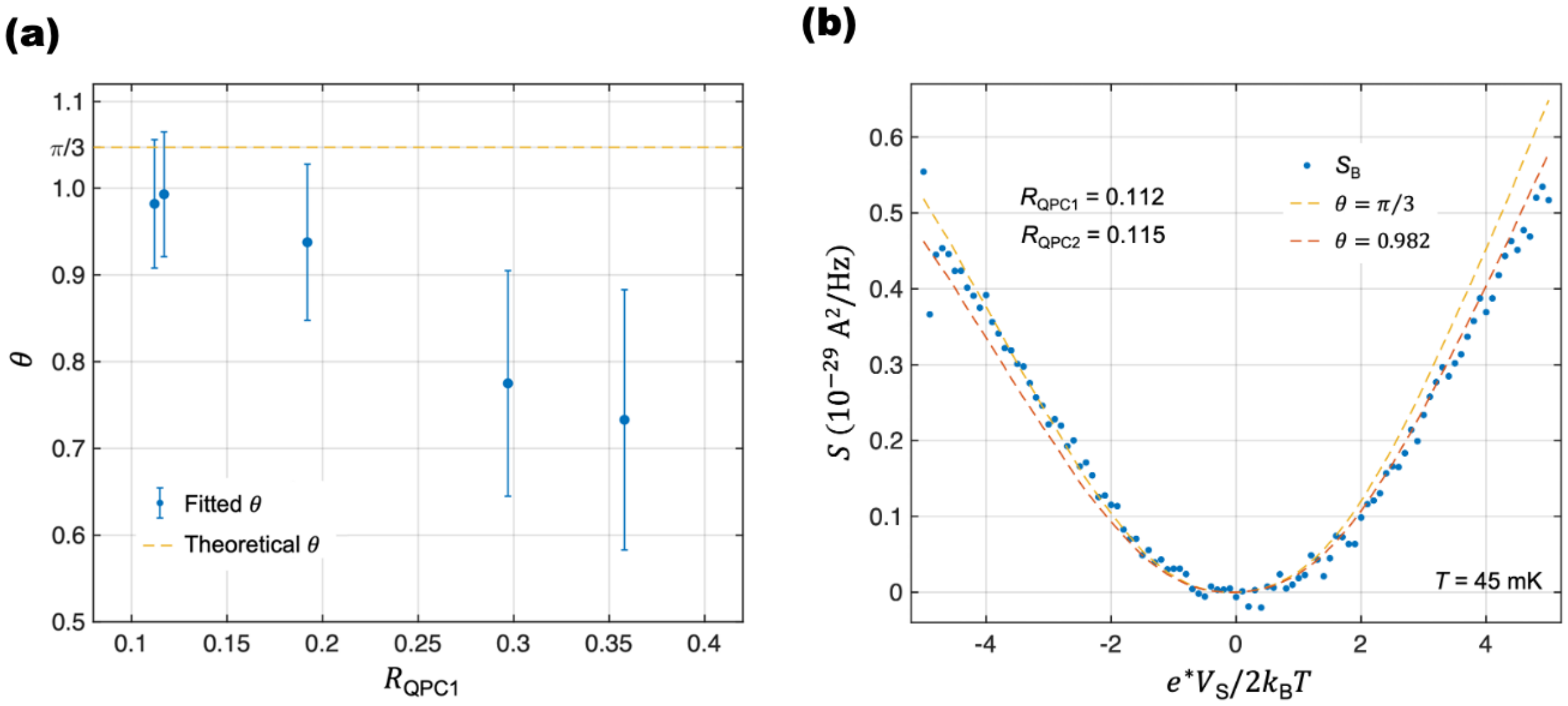

**FIG. S5: Fitting of exchange phase $\theta$. (a)** The result of fitting the data to the phenomenological model of Eq. (2) and Eq. (7) in the main text, with $\theta$ as the fitting parameter and $\delta = 1/3$ (blue dots), is plotted against the averaged $R_{\mathrm{QPC1}}$. From the most dilute to the least dilute case, each data point corresponds to Fig. 3 & Fig. 4(a-d) respectively. The theoretical value of $\theta = \pi/3$ is indicated by the yellow dashed line. **(b)** The curve with the best fitted value of $\theta$ (red dashed line) is shown with the experimental data (blue dots) for the most dilute case, corresponding to Fig. 3. For comparison, the theoretical curve with the ideal value of $\theta = \pi/3$(yellow dashed line) is shown together. Note that the curve is slightly different from the one in Fig. 3, as we here do not include the contribution of the trivial partition process and the finite temperature effects to $\mathcal{F}_{\mathrm{dilute}}$.

## SIII. THEORY OF THE ANOMALOUS PARTITION NOISE

### A. Fano factor $\mathcal{F}_{\mathrm{dilute}}$

We provide the theory of the Fano factor $\mathcal{F}_{\mathrm{dilute}}$ at sufficiently small $R_{\mathrm{QPC2}}$ and $e^*V_{\mathrm{S}} \gg k_{\mathrm{B}}T$. In Ref. [S1], the non-equilibrium correlator of anyon tunneling at QPC2 was derived for the dilute limit of $R_{\mathrm{QPC1}} \ll 1$.

We first restate the result for Abelian anyons [S1]. The tunneling operator at QPC2 is expressed as $\mathcal{T}(t) = \gamma_2\psi_3^\dagger(0,t)\psi_2(0,t)$, where $\psi_i(x,t)$ is the anyon annihilation operator on Edge$\boldsymbol{i}$ at position $x$ and time $t$, and $\gamma_2$ is the tunneling strength at QPC2. For simplicity, the position of QPC2 is chosen as $x = 0$ on both Edge2 and Edge3. The non-equilibrium correlator of the tunneling operator $C_{\mathrm{neq}}(t_1,t_2) \equiv \langle \mathcal{T}(t_1)\mathcal{T}^\dagger(t_2)\rangle_{\mathrm{neq}}$ in the presence of the dilute anyon beam is related to the equilibrium correlator $C_{\mathrm{eq}}(t_1,t_2) \equiv \langle \mathcal{T}(t_1)\mathcal{T}^\dagger(t_2)\rangle_{\mathrm{eq}}$ in the absence of the beam,

$$C_{\mathrm{neq}}(t_1,t_2) = e^{-\frac{I_{\mathrm{QPC1}}}{e^*}\left(e^{i2\theta\,\mathrm{sign}(t_2-t_1)}-1\right)|t_2-t_1|}C_{\mathrm{eq}}(t_1,t_2) + \text{subleading terms.} \tag{S1}$$

This was derived with the firm theoretical ground based on the conformal field theory or the bosonization (the chiral Luttinger liquid (CLL) theory) for FQH edge channels, combined with the Keldysh perturbation theory for arbitrary orders of anyon tunneling at QPC1.

The multiplicative factor, the non-equilibrium part of the expression of $C_{\mathrm{neq}}(t_1,t_2)$, is a consequence of time-domain anyon braiding. We found that the factor equals the average of the braiding phase $e^{2ik\theta}$ accumulated when the time-domain loop of a thermally excited anyon braids with $k$ anyons of the dilute beam arriving at QPC2 in the time interval $t_2 - t_1$ ($\gg h/e^*\mathrm{V_S}$),

$$\langle e^{2ik\theta}\rangle_{\mathrm{Poissonian}} = \sum_{k=0}^{\infty} Q_k e^{2ik\theta} = e^{-\frac{I_{\mathrm{QPC1}}}{e^*}\left(e^{2i\theta}-1\right)(t_2-t_1)}. \tag{S2}$$

The probability $Q_k$ of the event of $k$ anyons arriving at QPC2 in the interval $t_2 - t_1$ follows the Poissonian distribution $Q_k = \frac{m^k}{k!}e^{-m}$, and $m = I_{\mathrm{QPC1}}(t_2 - t_1)/e^*$ is the average number of anyons arriving at QPC2 in the interval $t_2 - t_1$. The Poisson distribution is natural, since anyons of the dilute beam is generated by tunneling from Edge1 to Edge2 at QPC1 in the regime of $R_{\mathrm{QPC1}} \ll 1$.

For a less dilute beam with relatively large $R_{\mathrm{QPC1}}$, yet small enough for the anyon tunneling, it is natural to expect that the time distribution of anyons of the beam follows a binomial distribution, instead of the Poissonian distribution. Hence, in our phenomenological theory, we replace the multiplicative factor $\langle e^{2ik\theta}\rangle_{\mathrm{Poissonian}}$ by the average braiding phase $\langle e^{2ik\theta}\rangle_{\mathrm{binomial}}$ over the binomial distribution $P_k$ of the number $k$,

$$\langle e^{2ik\theta}\rangle_{\mathrm{binomial}} = \sum_{k=0}^{n} P_k e^{2ik\theta} = \left(1 - R_{\mathrm{QPC1}} + R_{\mathrm{QPC1}}e^{2i\theta}\right)^{I_{\mathrm{S}}(t_2-t_1)/e^*} \tag{S3}$$

where $P_k = \frac{n!}{k!(n-k)!}R_{\mathrm{QPC1}}^k\left(1 - R_{\mathrm{QPC1}}\right)^{n-k}$, and $n = I_{\mathrm{S}}(t_2 - t_1)/e^*$ is the number of anyons impinging at QPC1 on Edge1 in the time interval $t_2 - t_1$ ($\gg h/e^*\mathrm{V_S}$). Using the factor, we write the non-equilibrium correlator,

$$C_{\mathrm{neq}}(t_1,t_2) = \left(1 - R_{\mathrm{QPC1}} + R_{\mathrm{QPC1}}e^{2i\theta\,\mathrm{sign}(t_2-t_1)}\right)^{I_{\mathrm{S}}|t_2-t_1|/e^*}C_{\mathrm{eq}}(t_1,t_2) + \text{subleading terms.} \tag{S4}$$

This expression is also applicable to the case of $t_2 < t_1$, in which the braiding direction of the time-domain loop is opposite to the $t_2 > t_1$ case. Eq. (S4) reduces to the previous results of Eq. (S1) for $R_{\mathrm{QPC1}} \ll 1$. This equation is valid for the long time regime $|t_2 - t_1| \gg h/e^*\mathrm{V_S}$, where the spatial width of the wave packet of anyons in the dilute beam is sufficiently narrow so that the time-domain braiding is well-defined. The sub-leading terms describe the trivial partition process and become important in the short time regime of $|t_2 - t_1| \simeq h/e^*\mathrm{V_S}$.

Once the non-equilibrium correlator is obtained, it is straightforward to calculate the tunneling rates $W_{2\to3}$ and $W_{3\to2}$,

$$W_{2\to3} = \int_{-\infty}^{\infty} dt \langle \mathcal{T}^\dagger(0)\mathcal{T}(t)\rangle_{\text{neq}}\,, \quad W_{3\to2} = \int_{-\infty}^{\infty} dt \langle \mathcal{T}(t)\mathcal{T}^\dagger(0)\rangle_{\text{neq}}\,. \tag{S5}$$

We first compute the contribution from the long time $|t_2 - t_1| \gg h/e^*V_S$ described by the time-domain braiding process,

$$\begin{aligned} W_{2\to3}^{\text{braid}} &= 2\frac{|\gamma_2|^2}{\hbar^2}\Gamma(1-2\delta)\text{Re}\left[e^{i\pi\delta}\left(-\frac{I_S}{e^*}\log\left(1+R_{\text{QPC1}}\left(e^{-2i\theta}-1\right)\right)\right)^{2\delta-1}\right], \\ W_{3\to2}^{\text{braid}} &= 2\frac{|\gamma_2|^2}{\hbar^2}\Gamma(1-2\delta)\text{Re}\left[e^{i\pi\delta}\left(-\frac{I_S}{e^*}\log\left(1+R_{\text{QPC1}}\left(e^{2i\theta}-1\right)\right)\right)^{2\delta-1}\right]. \end{aligned} \tag{S6}$$

Note that the only difference between the two rates is the braiding phase factor, $e^{-2i\theta} \leftrightarrow e^{2i\theta}$. This is explained by the fact that the the braiding direction of the time-domain loop is opposite between the processes of the two rates, particle tunneling from Edge2 to Edge3 for $W_{2\to3}^{\text{braid}}$ and hole tunneling from Edge2 to Edge3 for $W_{3\to2}^{\text{braid}}$. The time-domain braiding process contributes to the tunneling current and noise across QPC2 as $I_{\text{QPC2}}^{\text{braid}} = e^*\left(W_{2\to3}^{\text{braid}} - W_{3\to2}^{\text{braid}}\right)$ and $S_{\text{QPC2}}^{\text{braid}} = 2(e^*)^2\left(W_{2\to3}^{\text{braid}} + W_{3\to2}^{\text{braid}}\right)$,

$$\begin{aligned} I_{\text{QPC2}}^{\text{braid}} &= -4\frac{e^*}{\hbar^2}|\gamma_2|^2\Gamma(1-2\delta)\sin\pi\delta\,\text{Im}\left[\left(-\frac{I_S}{e^*}\log\left(1+R_{\text{QPC1}}\left(e^{-2i\theta}-1\right)\right)\right)^{2\delta-1}\right], \\ S_{\text{QPC2}}^{\text{braid}} &= 8\frac{e^{*2}}{\hbar^2}|\gamma_2|^2\Gamma(1-2\delta)\cos\pi\delta\,\text{Re}\left[\left(-\frac{I_S}{e^*}\log\left(1+R_{\text{QPC1}}\left(e^{-2i\theta}-1\right)\right)\right)^{2\delta-1}\right]. \end{aligned} \tag{S7}$$

If only the time-domain braiding determines the current and noise, the Fano factor $\mathcal{F}_{\text{dilute}}$ is written as,

$$\mathcal{F}_{\text{dilute}} \simeq \frac{S_{\text{QPC2}}^{\text{braid}}}{2e^* I_{\text{QPC2}}^{\text{braid}}} = -\cot\pi\delta\frac{\text{Re}\left[\left(-\log\left(1+R_{\text{QPC1}}\left(e^{-2i\theta}-1\right)\right)\right)^{2\delta-1}\right]}{\text{Im}\left[\left(-\log\left(1+R_{\text{QPC1}}\left(e^{-2i\theta}-1\right)\right)\right)^{2\delta-1}\right]}. \tag{S8}$$

The dependence of $\mathcal{F}_{\text{dilute}}$ on the diluteness $R_{\text{QPC1}}$ is plotted as the blue curve in Fig. S6. The Fano factor approaches to $\mathcal{F}_{\text{dilute}} \simeq 3.27$ in the Poissonian limit $R_{\text{QPC1}} \ll 1$, and decreases as the beam becomes less dilute.

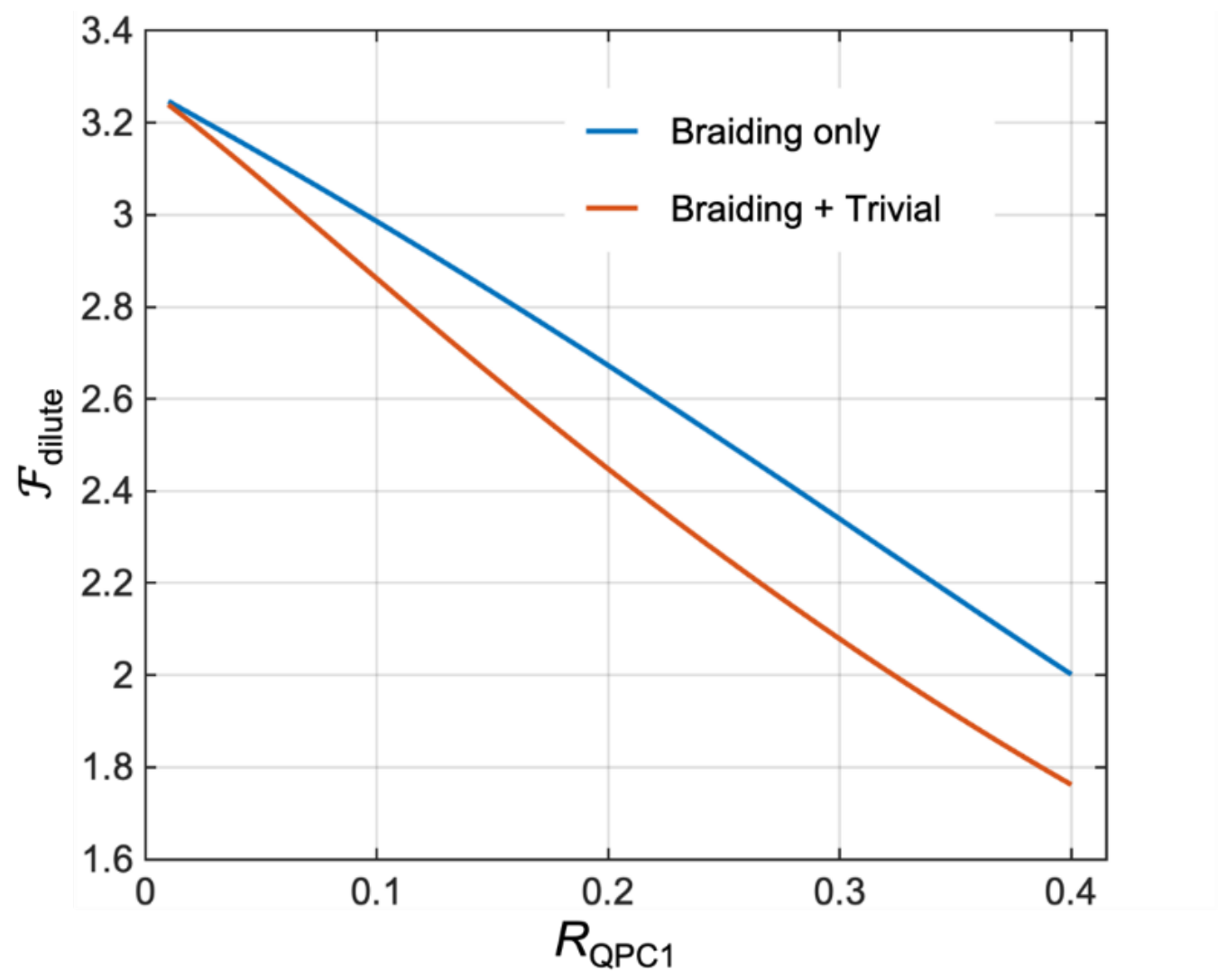

**FIG. S6: Dependence of Fano factor $\mathcal{F}_{\text{dilute}}$ on the diluteness $R_{\text{QPC1}}$.** The blue curve shows the Fano factor computed only with the time-domain braiding process, while the red curve shows the Fano factor contributed from both the time-domain braiding process and the trivial partition process.

There is also the trivial partition process, in which an anyon of the dilute beam directly tunnels at QPC2. This process occurs with the short time of $|t_2 - t_1| \simeq h/e^* V_S$. It is sub-dominant in contributing to the current and noise at QPC2, and described by the sub-leading terms in Eq. (S1) and Eq. (S4),

$$\text{the sub-leading terms of Eq. (S1) and Eq. (S4)} \simeq \frac{e}{e^*}\frac{\Gamma(2\delta)}{\Gamma(\delta)^2} R_{\mathrm{QPC1}} e^{-ie^* V_S (t_1 - t_2)/\hbar} C_{\mathrm{eq}}(t_1, t_2)\,. \tag{S9}$$

Using this, the contribution of the trivial partition process to the current and noise at QPC2 is obtained,

$$I_{\mathrm{QPC2}}^{\mathrm{trivial}} = e^* W_{2\to 3}^{\mathrm{trivial}} = R_{\mathrm{QPC1}} \times \frac{2\pi e|\gamma_2|^2}{\hbar^2 \Gamma(\delta)^2}\left(\frac{2\pi I_S}{e}\right)^{2\delta-1}, \quad S_{\mathrm{QPC2}}^{\mathrm{trivial}} = 2e^* I_{\mathrm{QPC2}}^{\mathrm{trivial}}\,. \tag{S10}$$

For the dilute limit $R_{\mathrm{QPC1}} \ll 1$, the contribution of the trivial process is sub-dominant compared to that of the braiding process,

$$\frac{\text{contribution of the trivial partition}}{\text{contribution of the time-domain braiding}} \propto R_{\mathrm{QPC1}}^{2-2\delta}. \tag{S11}$$

We note that if there were only the trivial partition process, the Fano factor has the value of $\mathcal{F}_{\mathrm{dilute}} = S_{\mathrm{QPC2}}^{\mathrm{trivial}} /(2e^* I_{\mathrm{QPC2}}^{\mathrm{trivial}}) = 1$ [see Eq. (S10)], as discussed in the main text.

The total current and noise at QPC2 are $I_{\mathrm{QPC2}} = I_{\mathrm{QPC2}}^{\mathrm{braid}} + I_{\mathrm{QPC2}}^{\mathrm{trivial}}$ and $S_{\mathrm{QPC2}} = S_{\mathrm{QPC2}}^{\mathrm{braid}} + S_{\mathrm{QPC2}}^{\mathrm{trivial}}$, where both the time-domain braiding process and the trivial partition process are taken into account. The full Fano factor is,

$$\mathcal{F}_{\mathrm{dilute}} = \frac{S_{\mathrm{QPC2}}}{2e^* I_{\mathrm{QPC2}}} = \frac{S_{\mathrm{QPC2}}^{\mathrm{braid}} + S_{\mathrm{QPC2}}^{\mathrm{trivial}}}{2e^*\left(I_{\mathrm{QPC2}}^{\mathrm{braid}} + I_{\mathrm{QPC2}}^{\mathrm{trivial}}\right)}. \tag{S12}$$

The dependence of the full Fano factor on $R_{\mathrm{QPC1}}$ is shown as the red curve in Fig. S6. As the contribution of the trivial partition process becomes larger (yet smaller than that of the braiding process) for larger $R_{\mathrm{QPC1}}$, the Fano factor further decreases.

## B. Phenomenological extension in Eq. (2)

In the last sub-section we have derived $S_{\mathrm{QPC2}} = \mathcal{F}_{\mathrm{dilute}} \times 2e^* I_{\mathrm{QPC2}}$ for high voltage $e^* V_S \gg k_B T$ and small QPC2 reflection $R_{\mathrm{QPC2}} \ll 1$. We now phenomenologically extend it to the form in Eq. (2) of the main text, to compare the result with our experimental data in a wider range of the parameters. We restate the equation,

$$S_{\mathrm{QPC2}} = \mathcal{F}_{\mathrm{dilute}} \times 2e^* I_{\mathrm{QPC1}} R_{\mathrm{QPC2}}\left(1 - R_{\mathrm{QPC2}}\right)\left[\coth\left(\frac{e^* V_S}{2k_B T}\right) - \frac{2k_B T}{e^* V_S}\right]. \tag{S13}$$

We here provide the rationale behind the extension.

Firstly, $[\coth(e^* V_S/2k_B T) - 2k_B T/e^* V_S]$ is introduced to describe a parameter range of relatively small values of voltage $V_S$. The factor $\coth(e^* V_S/2k_B T)$ comes from the hole-like anyon injection process at the QPC1 [S2]. It can be considered that the hole-like anyons are incoming from the source to QPC1 with a rate $I_S \exp(-e^* V_S/k_B T)/e^*$, which is exponentially suppressed in comparison with the particle-like anyon injection. The hole-like anyon injection affects both the time-domain braiding process and the trivial partition process. In the time-domain braiding process, when a hole-like anyon is injected at QPC1 to the dilute beam flowing along Edge2, the time-domain loop of a thermally excited anyon at QPC2 can braid the hole-like anyon, giving rise to the braiding phase factors of $e^{\pm 2i\theta}$, instead of the factors $e^{\mp 2i\theta}$ of the case of the particle-like anyon injection in Eq. (S4). As a result, the first term of the non-equilibrium correlator in Eq. (S4) has an additional multiplicative factor of $\left(1 - R_{\mathrm{QPC1}} + R_{\mathrm{QPC1}} e^{-2i\theta \operatorname{sign}(t_2 - t_1)}\right)^{\frac{I_S}{e^*}\exp\left(-\frac{e^* V_S}{k_B T}\right)|t_2 - t_1|}$ coming from the hole-like

anyon injection. Then the current and noise at QPC2 are modified accordingly. On the other hand, in the trivial partition process, the hole-like anyon injection modifies the current and noise at QPC2 as $I_{\mathrm{QPC2}}^{\mathrm{trivial}} \to I_{\mathrm{QPC2}}^{\mathrm{trivial}}\left(1-\exp\left(-\frac{e^*V_\mathrm{S}}{k_\mathrm{B}T}\right)\right)$ and $S_{\mathrm{QPC2}}^{\mathrm{trivial}} \to S_{\mathrm{QPC2}}^{\mathrm{trivial}}\left(1+\exp\left(-\frac{e^*V_\mathrm{S}}{k_\mathrm{B}T}\right)\right)$. The Fano factor $\mathcal{F}_{\mathrm{dilute}}$ is then calculated by $\mathcal{F}_{\mathrm{dilute}} = \frac{S_{\mathrm{QPC2}}}{2e^* I_{\mathrm{QPC2}} \coth(e^*V_\mathrm{S}/2k_\mathrm{B}T)}$. The last term of Eq. (S13) proportional to $-2k_\mathrm{B}T/e^*V_\mathrm{S}$ is introduced to make the excess noise to vanish at the zero bias of $V_\mathrm{S}=0$. Note that all the temperature dependence is introduced from the detailed balance principle, while we used the zero-temperature correlator of the CLL theory in the calculation of $\mathcal{F}_{\mathrm{dilute}}$. See also Supplementary Note IV. This phenomenological treatment is in analogy to the full beam case [S2], which remedies the power-law temperature dependence of the CLL theory that disagrees with experiments. In Fig. S7, we plot the calculated value of $\mathcal{F}_{\mathrm{dilute}}$ as a function of $e^*V_\mathrm{S}/2k_\mathrm{B}T$. For the calculation, the reflection probability $R_{\mathrm{QPC1}}$ in Fig. S4 is used.

Secondly, we did the substitution of $R_{\mathrm{QPC2}}$ to $R_{\mathrm{QPC2}}(1-R_{\mathrm{QPC2}})$ to obtain Eq. (S13). With this substitution, a parameter range of relatively large values of $R_{\mathrm{QPC2}}$ is described by Eq. (S13). This is done in the same spirit with the phenomenological expression of Eq.(1) for the partition of a full beam, where the substitution of $R_{\mathrm{QPC1}}$ to $R_{\mathrm{QPC1}}(1-R_{\mathrm{QPC1}})$ has been performed [S3] to have comparison between experimental data and the phenomenological expression in determination of fractional charges by shot noise, going beyond the parameter regime of the CLL theory. Excellent agreement between the phenomenological expression in Eq. (S13) (namely Eq. (2) of the main text) with our experimental data is found as shown in Figs. 3 & 4 of the main text.

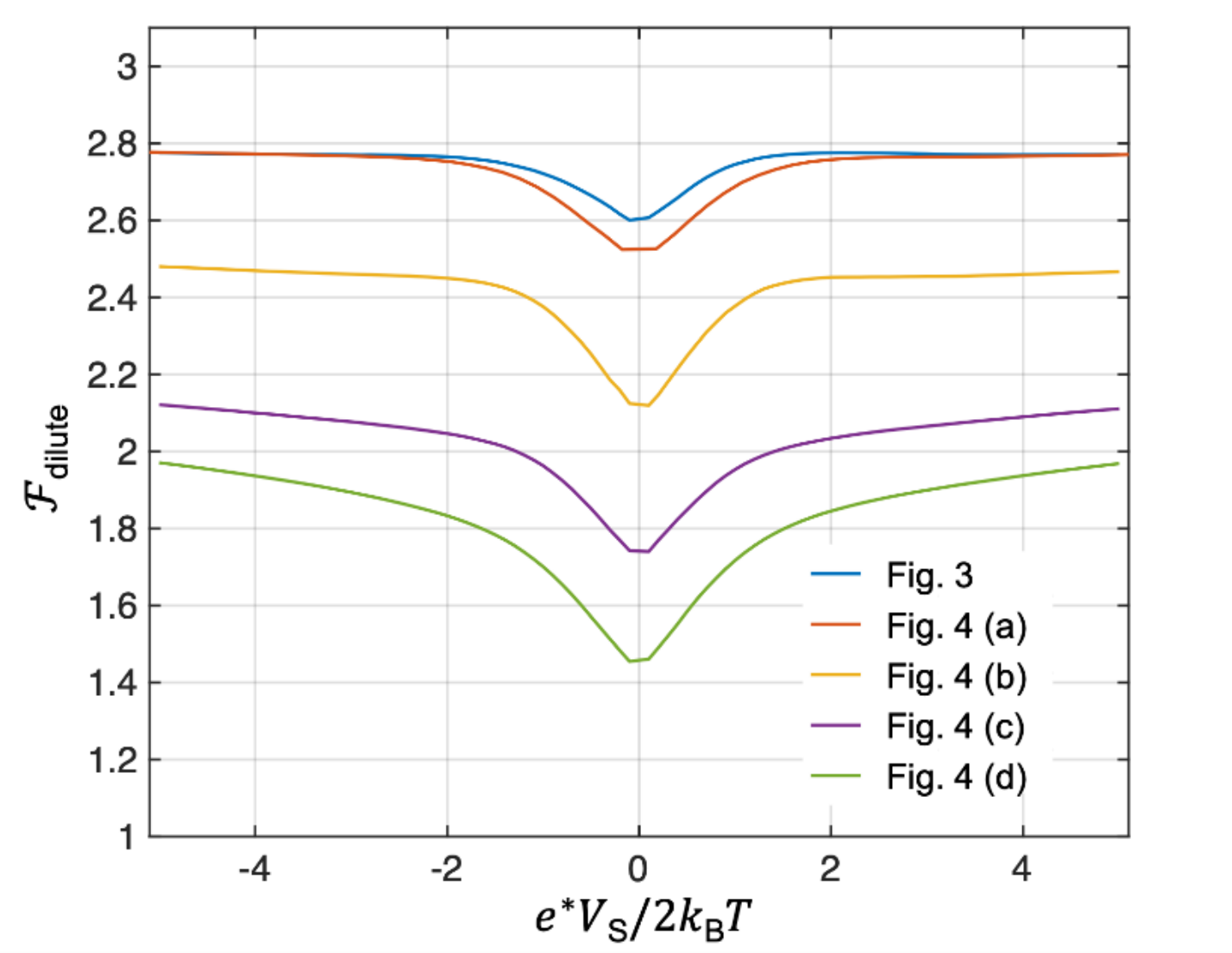

**FIG. S7: Calculated values of $\mathcal{F}_{\mathrm{dilute}}$, as a function of $e^*V_\mathrm{S}/2k_\mathrm{B}T$.** The values are used for drawing the theoretical curves in Figs. 3 and 4 of the main text. In the calculation of $\mathcal{F}_{\mathrm{dilute}}$, the experimental results of the reflection probability $R_{\mathrm{QPC1}}$ in Fig. S4.

### C. Dependence of Fano factor $\mathcal{F}_{\mathrm{dilute}}$ on the scaling dimension $\delta$

While the chiral Luttinger liquid theory predicts the power law behavior $R_{\mathrm{QPC}} \propto V^{2\delta-2}$ of the reflection probability $R_{\mathrm{QPC}}$ at a QPC with respect to a bias voltage $V$, this expected behavior has not been confirmed by experiments [S4]. As the Fano factor $\mathcal{F}_{\mathrm{dilute}}$ of our theory also depends on the scaling dimension $\delta$, it is in fact surprising that the excellent agreement between the theory and our experiment is found over a wide range of the parameters. To understand why, we investigate how $\mathcal{F}_{\mathrm{dilute}}$ varies as a function of the scaling dimension $\delta$. For simplicity, we concentrate on the high voltage regime of $e^*V_\mathrm{S} \gg k_\mathrm{B}T$.

For the Poissonian limit of $R_{\mathrm{QPC1}} \ll 1$, the time-domain braiding process dominates the trivial partition process, and the Fano factor is written concisely,

$$\mathcal{F}_{\text{dilute}} = -\cot\pi\delta\cot\left(\left(\frac{\pi}{2}-\theta\right)(2\delta-1)\right). \quad \text{(S14)}$$

As non-ideal effects at QPCs usually affect the scaling dimension $\delta$ to become larger than its ideal value 1/3 at $\nu$ = 1/3 [S5, S6], we explore how the Fano factor varies as $\delta$ increases from the ideal value. The result is shown as the blue curve in Fig. S8. As $\delta$ increases, the Fano factor decreases from the ideal value ($\simeq$ 3.27) at $\delta$ = 1/3 to 3 at $\delta$ = 1/2, and increases back to the original value at $\delta$ = 2/3. It shows that the Fano factor varies less than 10 % over the range of 1/3 < $\delta$ < 2/3. This may in part explain the excellent agreement between the theory and the experiment.

Next we take the realistic value of $R_{\text{QPC1}} = 0.1$, as in Fig. 3 of the main text. If we consider the time-domain braiding process only, $\mathcal{F}_{\text{dilute}}$ starts from 3.13 at $\delta$ = 1/3, reduces to 2.87 at $\delta$ = 1/2, and increases back to the original value 3.13 at $\delta$ = 2/3 (see the red curve in Fig. S8). Again, the variation of $\mathcal{F}_{\text{dilute}}$ is less than 10% over the range of $\delta$. The variation becomes bigger if we also include the trivial partition process. It starts from 3.08 at $\delta$ = 1/3 and decreases monotonically to 2.36 at $\delta$ = 2/3 (see the yellow curve in Fig. S8). In this case, the difference becomes about 20%. The relatively big variation is because the trivial process is less suppressed for larger $\delta$, as expected from the ratio $R_{\text{QPC1}}^{2-2\delta}$ of the contribution of the trivial process to that of the braiding process shown in Eq. (S11). Still, however, the variation is not that strong compared to the variation range of $\delta$.

Nevertheless, our transmission data are nearly flat, corresponding to $\delta$ = 1 (Supplementary Note SII C). With $\delta$ = 1, the Fano factor in Eq. (S14) diverges and cannot explain our experimental results. This suggests to revisit the long-time issue of whether and how the scaling dimension can be obtained from experimental data of the voltage dependence of QPC transmission. For example, the QPC model Hamiltonian used in the chiral Luttinger liquid theory for the prediction of the voltage dependence (the power-law behavior $R_{\text{QPC}} \propto V^{2\delta-2}$ of the QPC reflection probability $R_{\text{QPC}}$ on a voltage $V$ ) might be too simplified; while the bare anyon-tunneling strength at a QPC has been assumed to be energy independent in the theory, it could be energy dependent in realistic situations, which distorts the predicted power-law behavior even when the scaling dimension remains around the ideal value. Or, measurements of other quantities might be useful for experimental identification of the scaling dimension (see, e.g., Ref. [S8]).

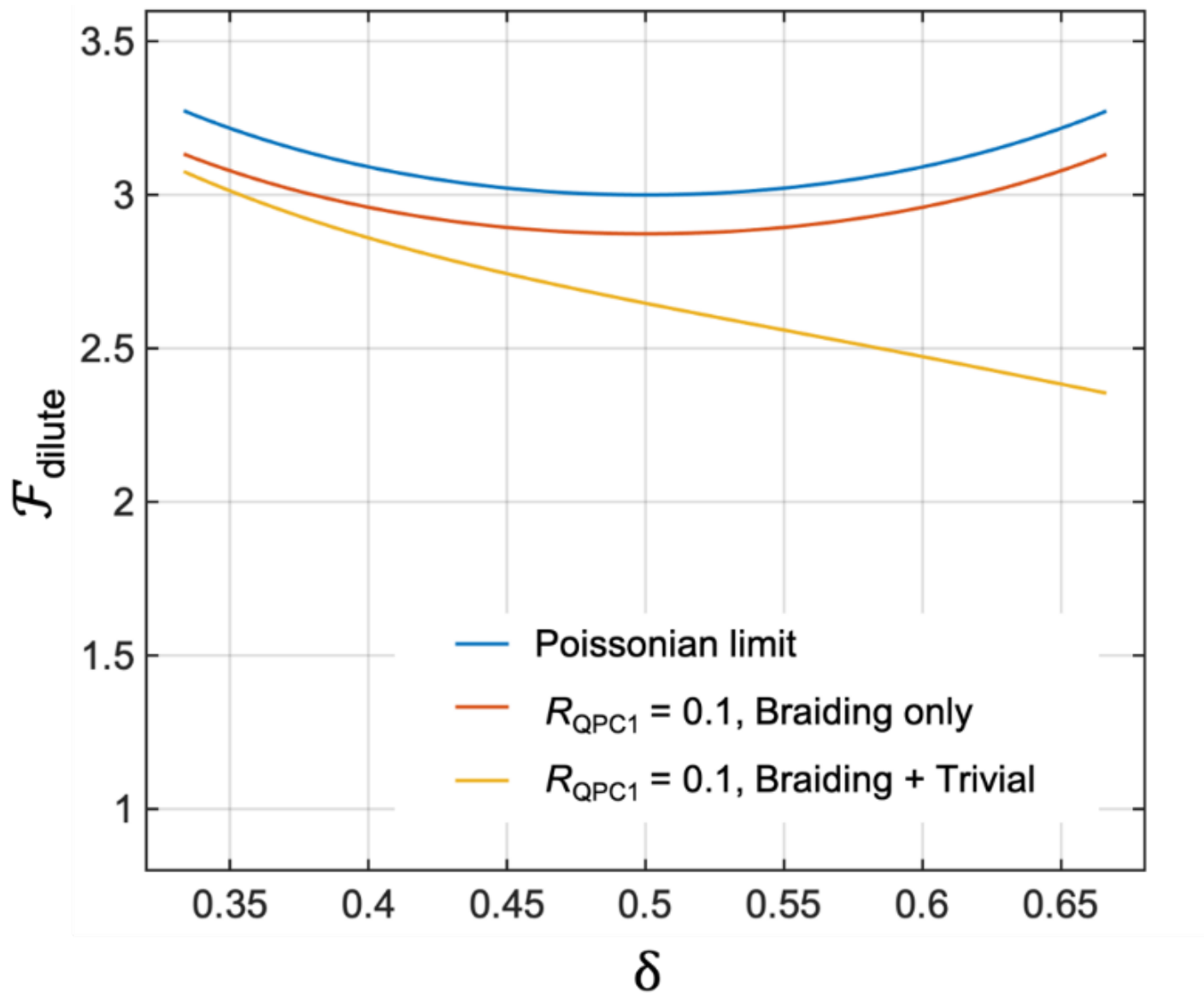

**FIG. S8: Dependence of Fano factor $\mathcal{F}_{\text{dilute}}$ on the scaling dimension $\delta$.** The blue curve is for the Poissonian limit, while the red and yellow curves are for $R_{\text{QPC1}} = 0.1$. In the red curve, only the time-domain braiding process is taken into account, while the yellow curve accounts both the braiding process and the trivial partition process.

### D. Illustration of the time-domain braiding process

We describe the time-domain braiding process in more details. In the process, anyons in the injected diluted beam propagate along Edge2 without tunneling at QPC2, while a particle-hole anyon pair is virtually or thermally excited at QPC2 and braids with some of the diluted anyons.

For illustration, we consider thermal equilibrium without injection of diluted anyons, with QPC2 tuned to the weak backscattering regime. The thermal fluctuations at QPC2 are described by the equilibrium correlator $C_{\text{eq}}(t_1, t_2) \equiv \langle \mathcal{T}(t_1)\mathcal{T}^\dagger(t_2)\rangle_{\text{eq}}$ of the anyon tunneling operator $\mathcal{T}(t)$ at QPC2 [introduced around Eq. (S1)]. The correlator is interpreted as the interference between two sub-processes $|\mathcal{T}(t_1)\rangle_{\text{eq}}$ and $|\mathcal{T}(t_2)\rangle_{\text{eq}}$. In the sub-process $|\mathcal{T}(t_1)\rangle_{\text{eq}}$, a particle-hole anyons pair is virtually or thermally excited at QPC2 at time $t_1$. Then, the *particle-like* anyon propagates along one edge, say Edge2, while the *hole-like* anyon propagates along the other edge, Edge3. In the other sub-process $|\mathcal{T}(t_2)\rangle_{\text{eq}}$, another pair is excited at QPC2 at time $t_2$, and the particle moves along Edge2 while the hole propagates along Edge3. The interference occurs by the overlap between $|\mathcal{T}(t_1)\rangle_{\text{eq}}$ and the time reversal of $|\mathcal{T}(t_2)\rangle_{\text{eq}}$, forming a time-domain loop at QPC2 as long as $t_2 - t_1 \lesssim \hbar/k_{\text{B}}T$ in which the particle tunnels to Edge2 at $t_1$, tunnels back to Edge3 at $t_2$, and recombines with the hole. The two sub-processes $|\mathcal{T}(t_1)\rangle_{\text{eq}}$ and $|\mathcal{T}(t_2)\rangle_{\text{eq}}$ correspond to those in Fig. S9(a) and (b) but without the diluted anyon (the red wave packet), and the blue dashed loop in Figure 2(b) of the main text provides a schematic view of the time-domain loop. The time-domain loop results in the known *equilibrium* noise of tunneling currents at QPC2. Mathematically, in the CLL theory, the tunneling current and noise are determined by an integral of the equilibrium correlator $C_{\text{eq}}(t_1, t_2) = \frac{1}{[\frac{\hbar}{\pi k_B T}\sin(\pi k_{\text{B}} T(a + i\,(t_1 - t_2))/\hbar]^{2\delta}}$ over $t_1$ and $t_2$ where $a \to 0^+$ is a short distance cutoff. The integral is governed by the domain of $t_2 - t_1 \lesssim \hbar/k_{\text{B}}T$, supporting the above interpretation.

We next consider the non-equilibrium by dilute anyons injection via QPC1, focusing on a simple case [the $k = 1$ event below Eq. (4) of the main text] where a single diluted anyon experiences braiding by the time-domain loop. It is the interference between the two sub-processes in Fig. S9 (a) and (b) [S1,S9]. In the sub-process (a), a particle-hole anyons pair is thermally excited at QPC2 at $t_1$ as in the equilibrium case. Later at time $t_0^{(a)}$, a diluted anyon passes QPC2 without partitioning. In the sub-process (b), the diluted anyon firstly passes QPC2 at time $t_0^{(b)}$, and then the particle-hole pair is excited at QPC2 at $t_2$. Due to the source voltage $V_S$, $t_0^{(a)}$and $t_0^{(b)}$ are in the range $|t_0^{(a)} - t_0^{(b)}| < \frac{\hbar}{e^* V_{\text{S}}}$, leading to $t_0^{(a)} \approx t_0^{(b)} \approx t_0$ at sufficiently large $V_S$. To summarize, in the sub-processes (a) and (b), the particle-hole pair is excited *before* and *after* the diluted anyon passes QPC2, respectively, i.e. $t_1 < t_0 < t_2$. The spatial order of the anyons differs between the sub-processes, resulting in the braiding phase in their interference (see Supplementary Video).

Analytically, the braiding phase stems from the double exchange between the diluted anyon and the thermal anyon. The interference is described by the non-equilibrium correlator $\langle \psi_2(-d, t_0^{(b)} - d/v)\mathcal{T}^\dagger(t_2)\mathcal{T}(t_1)\psi_2^\dagger(-d, t_0^{(a)} - d/v)\rangle$, where anyon creation operators $\psi_2^\dagger(-d, t_0^{(a)} - d/v)$ and $\psi_2^\dagger(-d, t_0^{(b)} - d/v)$ on Edge2 describe the injection of the diluted anyon via QPC1 in (a) and (b) (with the inter-QPC distance $d$, QPC1 placed at $x = -d$, the anyon velocity $v$). At sufficiently large $V_{\text{S}}$, we have $t_1 < t_0^{(a)} \simeq t_0^{(b)} < t_2$ and annihilate the operator $\psi_2^\dagger(-d, t_0^{(a)} - d/v)$ with $\psi_2(-d, t_0^{(b)} - d/v)$ in the correlator after bringing it in front of $\mathcal{T}(t_1)$ and $\mathcal{T}^\dagger(t_2)$ by operator exchanges. The double exchange results in the braiding phase $2\theta$, and the non-equilibrium correlator becomes equivalent with $e^{2i\theta}\langle \mathcal{T}^\dagger(t_2)\mathcal{T}(t_1)\rangle_{\text{eq}}$.

The braiding phase is well defined when the temporal uncertainty of the diluted anyon $|t_0^{(a)} - t_0^{(b)}| < \frac{\hbar}{e^* V_{\text{S}}}$ is small enough compared to $|t_1 - t_2|$, while $|t_1 - t_2|$ should be shorter than $\hbar/k_{\text{B}}T$ for the formation of the time-domain loop. These two conditions are satisfied in our interested regime of $e^* V_{\text{S}} \gg k_{\text{B}}T$. When $e^* V_{\text{S}}$ is not sufficiently large, the condition for the braiding $t_1 < t_0^{(a)} \simeq t_0^{(b)} < t_2$ is not satisfied, making the braiding phase blurred.

We remark that anyon braiding on a single edge channel (in 1D) is more abstract than real space braiding in two dimensions since the one-dimensional system has no room for an adiabatic circulation of one anyon around another. Analytically, an additional dimension for 1D braiding is provided by ordering of the anyon fields inside a correlator as discussed above, and the braiding on the edge is defined by double exchange of the anyon fields on the edge.

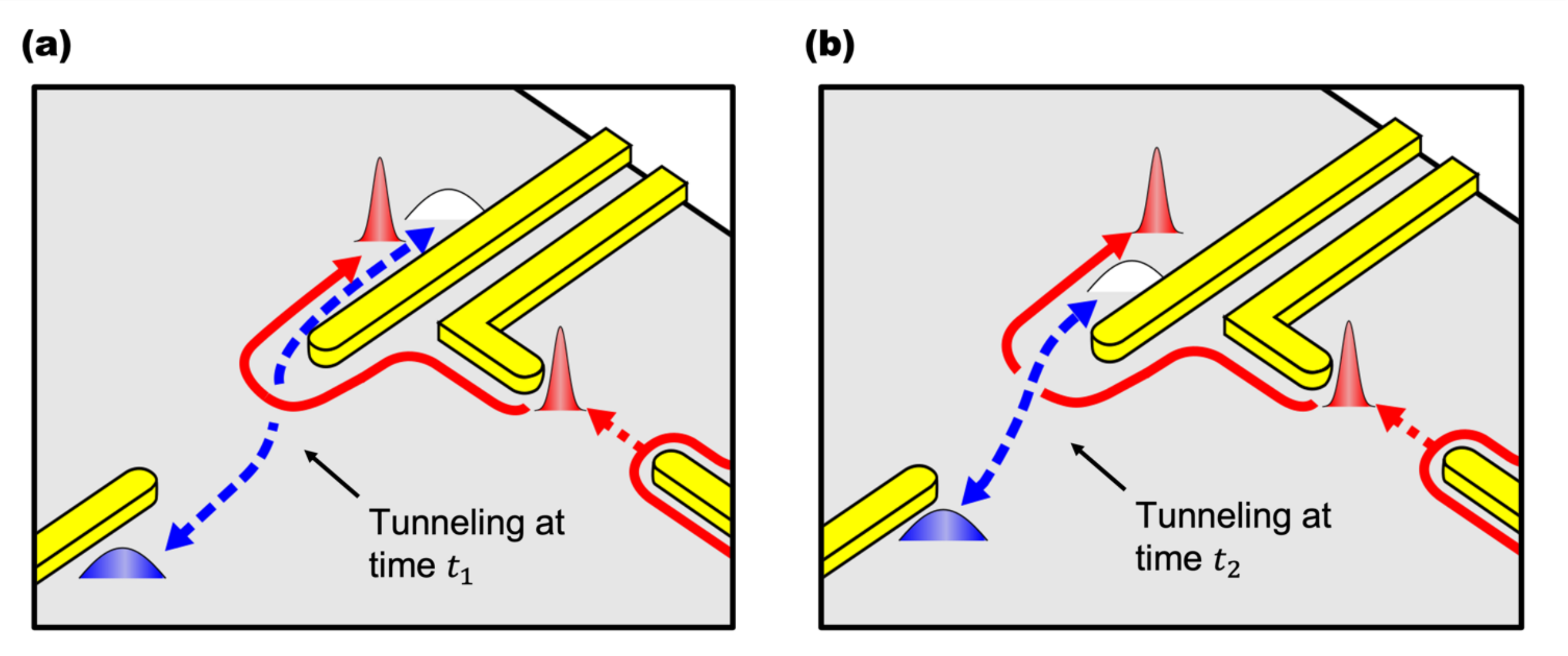

**FIG. S9: Illustration of the time-domain braiding process of the $k = 1$ event.** The braiding occurs by the interference of two sub-processes (a) and (b). In the sub-process (a), a particle-hole anyon pair (blue and white wave-packets) is thermally excited via tunneling at QPC2 at time $t_1$ *before* a diluted anyon (red wave-packet) passes QPC2. In (b), the particle-hole pair is excited at time $t_2$ *after* the diluted anyon passes QPC2. The interference of the two sub-processes forms a time-domain loop of the thermal anyons.

## SIV. COMPARISON BETWEEN $S_{\mathrm{QPC2}}$ AND $S_{\mathrm{B}}$.

In this supplementary note, we argue why our phenomenological treatment agrees well with the experiments, based on comparison among the experimental data, the excess noise $S_{\mathrm{QPC2}}$ in Eq. (2), which hybridizes the zero temperature CLL theory and its phenomenological extension to the finite temperature, and the excess noise $S_{\mathrm{B}}$ obtained from a finite temperature CLL theory.

For the purpose, we combine the finite temperature correlator of the CLL theory with the binomial extension. Then, the noise and current across QPC2 can be found as

$$I_{\mathrm{QPC2}}(I_{\mathrm{S}}, I_{\mathrm{ref}}) = -4\frac{e^*}{\hbar^2}|\gamma_2|^2\Gamma(1-2\delta)\sin\pi\delta\,(2\pi k_{\mathrm{B}}T/\hbar)^{2\delta-1}\mathrm{Im}\left[\frac{\Gamma\left(\frac{\mathcal{J}(I_{\mathrm{S}},I_{\mathrm{ref}})}{2\pi k_{\mathrm{B}}T/\hbar}+\delta\right)}{\Gamma\left(\frac{\mathcal{J}(I_{\mathrm{S}},I_{\mathrm{ref}})}{2\pi k_{\mathrm{B}}T/\hbar}+1-\delta\right)}\right],$$
$$S_{\mathrm{QPC2}}(I_{\mathrm{S}}, I_{\mathrm{ref}}) = 8\frac{e^{*2}}{\hbar^2}|\gamma_2|^2\Gamma(1-2\delta)\cos\pi\delta\,(2\pi k_{\mathrm{B}}T/\hbar)^{2\delta-1}\mathrm{Re}\left[\frac{\Gamma\left(\frac{\mathcal{J}(I_{\mathrm{S}},I_{\mathrm{ref}})}{2\pi k_{\mathrm{B}}T/\hbar}+\delta\right)}{\Gamma\left(\frac{\mathcal{J}(I_{\mathrm{S}},I_{\mathrm{ref}})}{2\pi k_{\mathrm{B}}T/\hbar}+1-\delta\right)}\right]. \tag{S15}$$

where $\mathcal{J}(I_{\mathrm{S}}, I_{\mathrm{ref}}) = -\frac{I_{\mathrm{S}}}{e^*}\log\left(1+R_{\mathrm{QPC1}}\left(e^{-2i\theta}-1\right)\right) - \frac{I_S}{e^*}\exp\left(-\frac{e^*V_{\mathrm{S}}}{k_{\mathrm{B}}T}\right)\log\left(1+R_{\mathrm{QPC1}}\left(e^{2i\theta}-1\right)\right) + i\frac{2\pi}{e}I_{\mathrm{ref}}$. Here, the first and second terms correspond to the braiding process from the particle-like anyons and hole-like anyons of the dilute beam respectively, and $I_{\mathrm{ref}} = -GV_3$ in the third term is the direct bias across the QPC2 [S1], with $G = \nu e^2/h$. $V_3$ is the small applied voltage to S2 while pinching QPC3 completely off.

The autocorrelation at the amplifier B and the autocorrelation across the QPC2 are related by,

$$S_{\mathrm{B}}(I_{\mathrm{S}}, I_{\mathrm{ref}}=0) = S_{\mathrm{QPC2}}(I_{\mathrm{S}}, I_{\mathrm{ref}}=0) - 4k_BTG\,\frac{\partial I_{\mathrm{QPC2}}(I_{\mathrm{S}}, I_{\mathrm{ref}})}{\partial I_{\mathrm{ref}}}\Big|_{I_{\mathrm{ref}}=0} + 4k_{\mathrm{B}}TG. \tag{S16}$$

Here, $S_{\mathrm{B}}(I_{\mathrm{S}}, I_{\mathrm{ref}}=0)$ and $S_{\mathrm{QPC2}}(I_S, I_{\mathrm{ref}}=0)$ are the bare noises, not the excess noises. The second term on the right hand side corresponds to the correlation between the tunneling current across QPC2 and the current flowing along Edge3. The third term is the Johnson-Nyquist noise of Edge3. While $S_{\mathrm{B}}(I_{\mathrm{S}}, I_{\mathrm{ref}}=0)$ increases as $|I_S|$ increases both theoretically and experimentally, the first and the second terms of the right hand side are theoretically expected to exhibit the maximal value for $I_{\mathrm{S}}=0$, reminiscent of the power-law behavior of the CLL theory. There was a theoretical proposal to extract the braiding statistics from the negative excess shot noise $S_{\mathrm{QPC2}}$ [S9]. However, the experimentally measured value of the second term shows the opposite behavior; while it is minimal at $I_{\mathrm{S}}=0$, it is relatively flat across the whole range of the bias. While it makes hard to observe negative excess shot noise of $S_{\mathrm{QPC2}}$, this leads the excess $S_{\mathrm{B}}$ and $S_{\mathrm{QPC2}}$ to almost coincide. This is why the expression of $S_{\mathrm{QPC2}}$ in Eq. (2) is directly used for analyzing the experimentally measured $S_{\mathrm{B}}$.

To calculate the excess noise $S_{\mathrm{B}}$, we rewrite $S_{\mathrm{B}}(I_S, I_{\mathrm{ref}}=0)$ in the following form

$$S_{\mathrm{B}}(I_{\mathrm{S}}, I_{\mathrm{ref}}=0) = 2e^*\left[\frac{S_{\mathrm{B}}(I_{\mathrm{S}}, I_{\mathrm{ref}}=0)}{2e^*I_{\mathrm{QPC2}}(I_S, I_{\mathrm{ref}}=0)\coth\left(\frac{e^*V_{\mathrm{S}}}{2k_{\mathrm{B}}T}\right)}\right]I_{\mathrm{QPC1}}R_{\mathrm{QPC2}}\left(1-R_{\mathrm{QPC2}}\right)\coth\left(\frac{e^*V_{\mathrm{S}}}{2k_BT}\right). \tag{S17}$$

The quantity in [...] is obtained by the theory and corresponds to the $\mathcal{F}_{\mathrm{dilute}}$ in Eq. (2), where non-universal effects in $S_{\mathrm{B}}$ and $I_{\mathrm{QPC2}}$ are expected to be cancelled in the ratio $S_{\mathrm{B}}/I_{\mathrm{QPC2}}$. We also did the phenomenological extension of $R_{\mathrm{QPC2}} \to R_{\mathrm{QPC2}}\left(1-R_{\mathrm{QPC2}}\right)$. For further calculation of $S_{\mathrm{B}}(I_{\mathrm{S}}, I_{\mathrm{ref}}=0)$ in Eq. (S17), we utilize the experimental values of $I_{\mathrm{QPC1}}$, $R_{\mathrm{QPC2}}$, $V_{\mathrm{S}}/T$. The excess noise $S_{\mathrm{B}}$ can then be calculated by subtracting the value at zero bias,.

$$S_{\mathrm{B}} = S_{\mathrm{B}}(I_{\mathrm{S}}, I_{\mathrm{ref}}=0) - S_{\mathrm{B}}(I_{\mathrm{S}}=0, I_{\mathrm{ref}}=0). \tag{S18}$$

Equations (S17) and (S18) is a phenomenological extension of the CLL theory for $S_{\mathrm{B}}$ , comparable with the extension for $S_{\mathrm{QPC2}}$ found in Eq. (2). In Fig. S10, we plot $S_{\mathrm{QPC2}}$ from Eq. (2) and $S_{\mathrm{B}}$ from Eq. (S18) together with the experimental data in Fig. 3. Almost perfect agreement among the three supports to use $S_{\mathrm{QPC2}}$ in Eq. (2) for the comparison with the experiments. It seems that the power-law finite temperature behavior of the CLL correlator, which is not observed in the experiment, is approximately canceled between the first and second term of Eq. (S16).

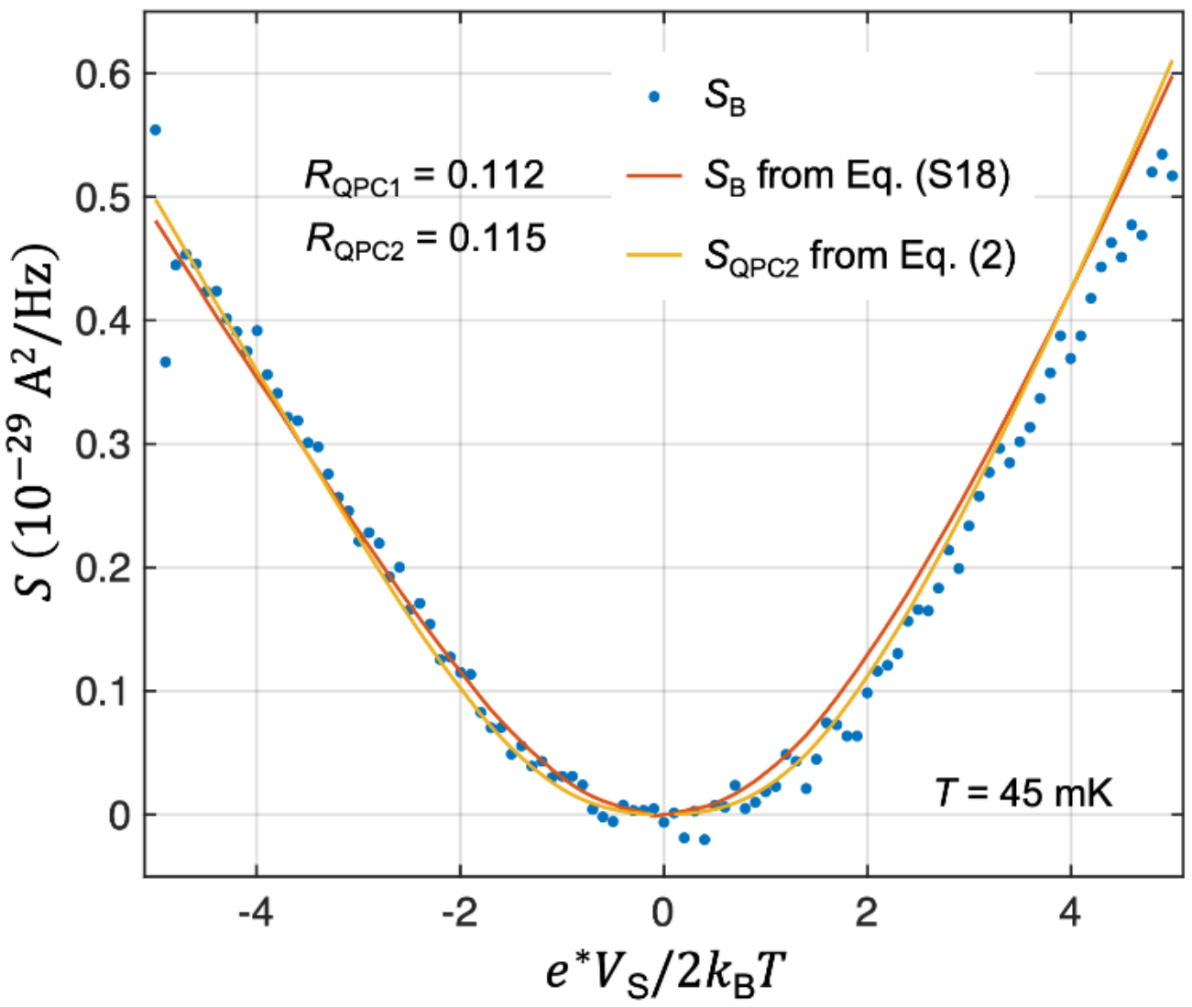

**FIG. S10**: **$S_B$ and $S_{QPC2}$**. The difference among $S_B$ from the experimental data (the data set in Fig. 3 of the main text), $S_B$ from Eq. (S18), and $S_{QPC2}$ from Eq. (2) is sufficiently small.

## SV. TWO-QPC EXPERIMENT AT FILLING FACTOR $\nu = 2/5$

In this supplementary note, we extend our study to the FQH regime at ν = 2/5. Its edge structure is composed of two (inner and outer) edge channels. The inter-QPC distance is 2 μm.

First, we performed a two-QPC experiment on the outer edge channel, making the inner edge channels fully reflected at the QPCs. In the same way with Fig. 1(b), we confirmed that the tunneling charge at QPC1 is $e^* = e/3$ [Fig. S11(a)], as expected. We observed that partitioning of diluted anyons of the fractional charge at QPC2 results in the Fano factor close to $\mathcal{F}_{\text{dilute}}$ = 3.27 as in ν = 1/3 [Fig. S11(b)]. The Fano factor agrees well with our theory based on the braiding angle $2\theta = 2\pi/3$ and the scaling dimension $\delta = 1/3$, supporting the time-domain braiding process also in the outer edge channel at ν = 2/5.

Next, we performed another two-QPC experiment on the inner edge channel, making the outer edge channels fully transmitted through the QPCs. The tunneling charge at QPC2 was found as $e^* = e/5$ from the shot noise measurement where a full beam impinges at QPC2 while QPC1 is pinched off [Fig. S12(a)]. Then, partitioning a dilute beam at QPC2, we obtained $\mathcal{F}_{\text{dilute}} \sim 1$ [Fig. S12(b)] in our measurement uncertainty (which suffers from the very weak spectral density, weaker than the ν = 1/3 case). The result implies that the trivial partition process is more substantial along the inner channel.

We compare the experimental result of partitioning the inner channel with existing theoretical models. There are several models for edges at ν = 2/5. In the model by Wen [S11] where large spatial separation between the inner and outer channels is considered, anyons with fractional charge $e^* = e/5$ have the braiding phase $2\theta = 6\pi/5$ and scaling dimension δ = 3/5. This model supports $\mathcal{F}_{\text{dilute}} \simeq -5.16$ when only the time-domain braiding process is considered and $\mathcal{F}_{\text{dilute}} \simeq 30$ when both the time-domain braiding and the trivial partition processes are considered with the measured value of $R_{\text{QPC1}} = 0.088$. Hence it is incompatible with our experiment. Another model proposed by Lopez and Fradkin [S12] predicts a downstream charge mode and non-propagating neutral modes. In this case, the braiding phase is solely from the propagating downstream charge mode, and it has the value of $2\theta = \pi/5$. And, the two $\delta$s appearing in Eq. (S14) become to have different values; the first one is 8/5 and the second is 1/10. This is because the non-propagating neutral mode contributes to the anyonic exchange phase at a QPC, but not to its tunneling exponent. The resulting Fano factor is $\mathcal{F}_{\text{dilute}} \simeq -0.2$, which cannot explain our experiment. On the other hand, Ferraro *et al.* [S13] modified the Lopez-Fradkin model. In their model, there is a downstream charge mode and upstream neutral modes. Then, while the most relevant tunneling charge at a QPC at low temperature is $e^* = 2e/5$ (which is described by $2\theta = 4\pi/5$ and $\delta = 2/5$), there is another quasiparticle tunneling of charge $e^* = e/5$ (described by $2\theta = \pi/5$ and $\delta = 8/5$). Since our experimental data support $e^* = e/5$ and it is expected that the quasiparticle of $e^* = e/5$ has larger bare QPC tunneling strength than that of $e^* = 2e/5$, we assumed that the anyon with $e^* = e/5$ dominates the QPC tunneling in our experiment. Our theory shows that this anyon results in the Fano factor $\mathcal{F}_{\text{dilute}} \simeq 1$, since the large scaling dimension $\delta = 8/5$ (> 1) makes the trivial partition process to dominate over the braiding process [See Eq. (S11)]. This may explain our experiment on partitioning the inner edge channel. There might be also a possibility that interactions between the channels give rise to decoherence effects in favor of the trivial partition process.

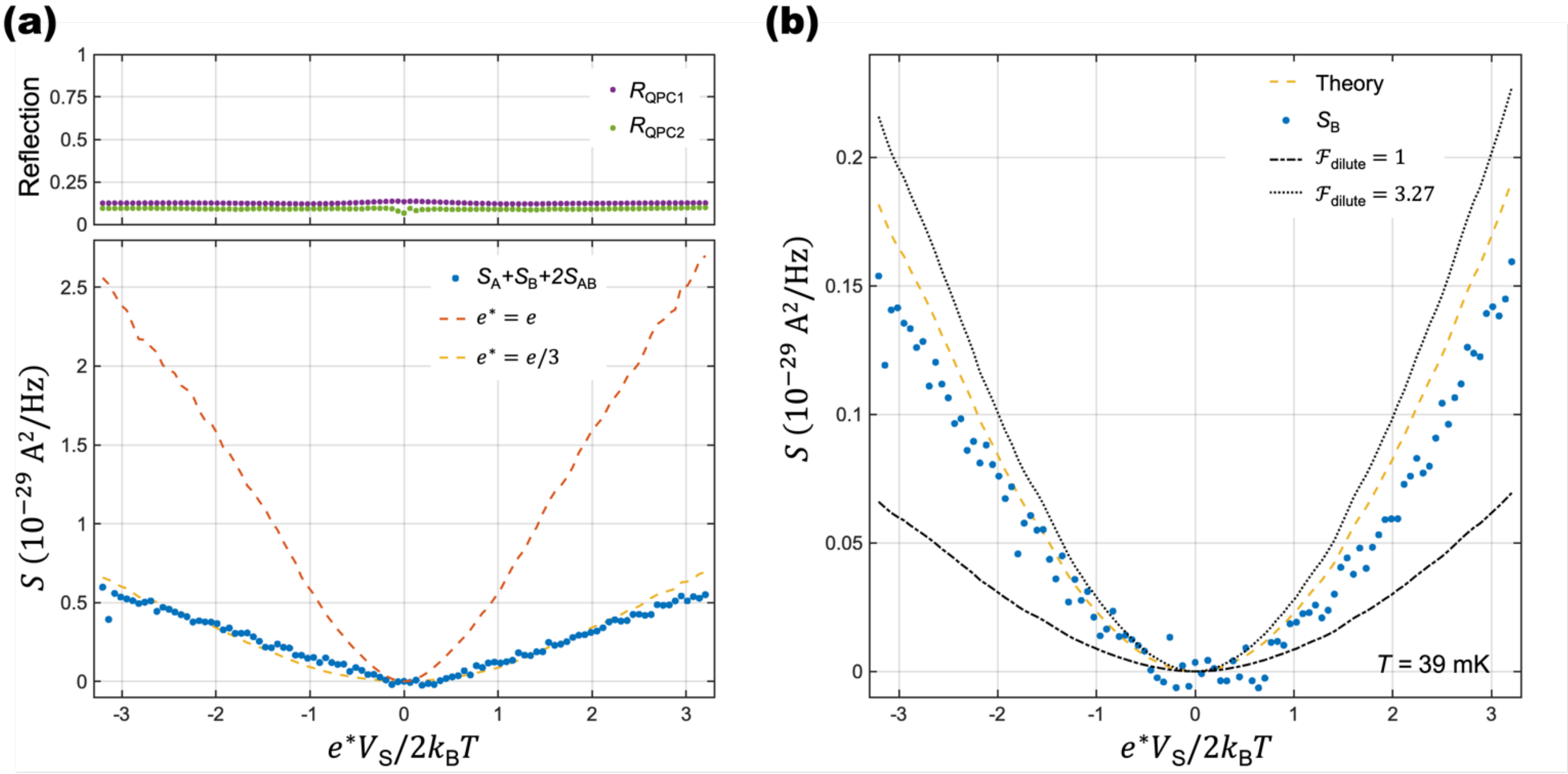

**FIG. S11: Noise measurement for ν = 2/5 outer edge. (a)** Upper panel: Reflection probabilities $R_{\mathrm{QPC1}}$ (purple dots) and $R_{\mathrm{QPC2}}$ (green dots). Lower panel: Measurement of tunneling charge at QPC1. Blue dots are $S_{\mathrm{A}} + S_{\mathrm{B}} + 2S_{\mathrm{AB}}$ noise calculation from measurement. Red dashed line and yellow dashed line are obtained from Eq. (1) in the main text, with $e^* = e$ and $e^* = e/3$ respectively. **(b)** Dilute beam impinges on QPC2, creating the excess AC. The reflection probabilities of the QPCs are the same as (a).

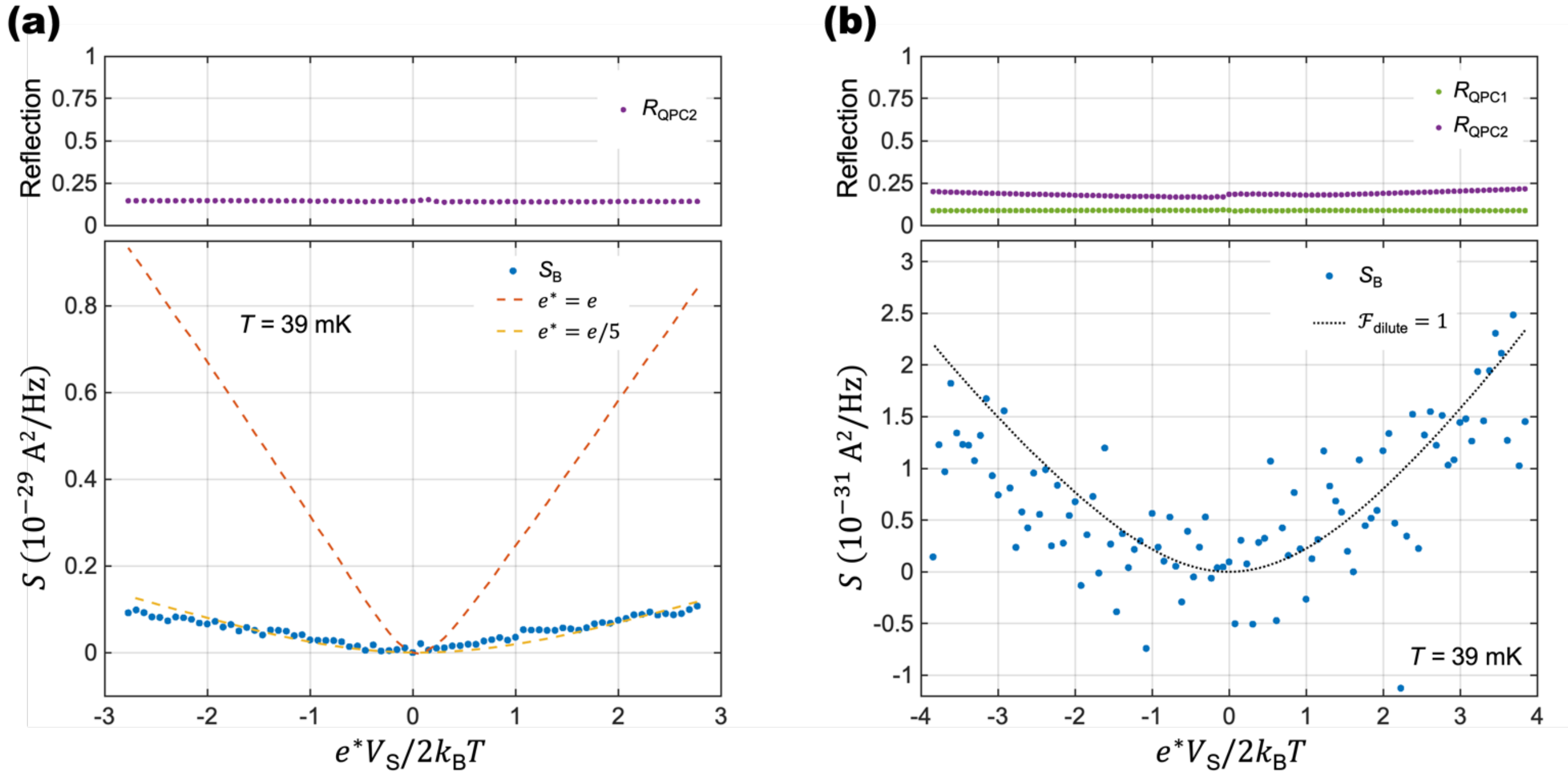

**FIG. S12: Noise measurement for ν = 2/5 inner edge. (a)** QPC1 is pinched off for injecting a full beam to QPC2. Upper panel: Reflection probability $R_{\mathrm{QPC2}}$ for the inner edge. Lower panel: Auto-correlation shot noise measurement results (blue dots) at 39 mK. Red dashed line and yellow dashed line are obtained from Eq. (1) in the main text, with $e^* = e$ and $e^* = e/5$ respectively. **(b)** Two-QPC experiment result. Upper panel: Reflection probability $R_{\mathrm{QPC1}}$ (green dots) and $R_{\mathrm{QPC2}}$ (purple dots). Lower panel: The measured excess noise $S_{\mathrm{B}}$ (blue dots). The black dotted line corresponds to Eq. (2) with $\mathcal{F}_{\mathrm{dilute}} = 1$.

## SVI. CROSS CORRELATION

Here, we show that our experimental data of the cross correlation (CC) $S_{\mathrm{AB}}$ of the two-QPC setup and an additional three-QPC setup are also in excellent agreement with our theory mainly based on the time-domain braiding process. For the two-QPC setup, the CC $S_{\mathrm{AB}}$ between amplifiers A and B is related to the AC noise of QPC2 by

$$S_{\mathrm{AB}} = -S_{\mathrm{QPC2}} + \frac{\partial I_{\mathrm{QPC2}}}{\partial I_{\mathrm{QPC1}}} S_{\mathrm{QPC1}}\,, \tag{S19}$$

where the second term corresponds to the correlation between the tunneling current $I_{\mathrm{QPC1}}$ at QPC1 and the tunneling current $I_{\mathrm{QPC2}}$ at QPC2 [S1, S7]. For comparison between our experimental data and the theory, we replace the differential reflection $\partial I_{\mathrm{QPC2}}/\partial I_{\mathrm{QPC1}}$ by its averaged value $R_{\mathrm{QPC2}}$. Then $S_{\mathrm{AB}}$ is obtained by using $S_{\mathrm{QPC2}}$ in Eq. (S13) and $S_{\mathrm{QPC1}}$ in Eq.(1) of the main text. This theoretical result is in good agreement with our experimental data [Fig. S13(a)]. For comparison, we also cite the free fermion results from the Landauer-Büttiker formalism, which corresponds to the trivial partition,

$$S_{\mathrm{AB}} = -2e^* I_{\mathrm{S}} R_{\mathrm{QPC1}}^2 R_{\mathrm{QPC2}}(1 - R_{\mathrm{QPC2}}) \left[\coth\left(\frac{e^* V_{\mathrm{S}}}{2k_{\mathrm{B}}T}\right) - \frac{2k_{\mathrm{B}}T}{e^* V_{\mathrm{S}}}\right], \tag{S20}$$

and plot it in Fig. S13(a) as the red dashed line with $e^* = e/3$.

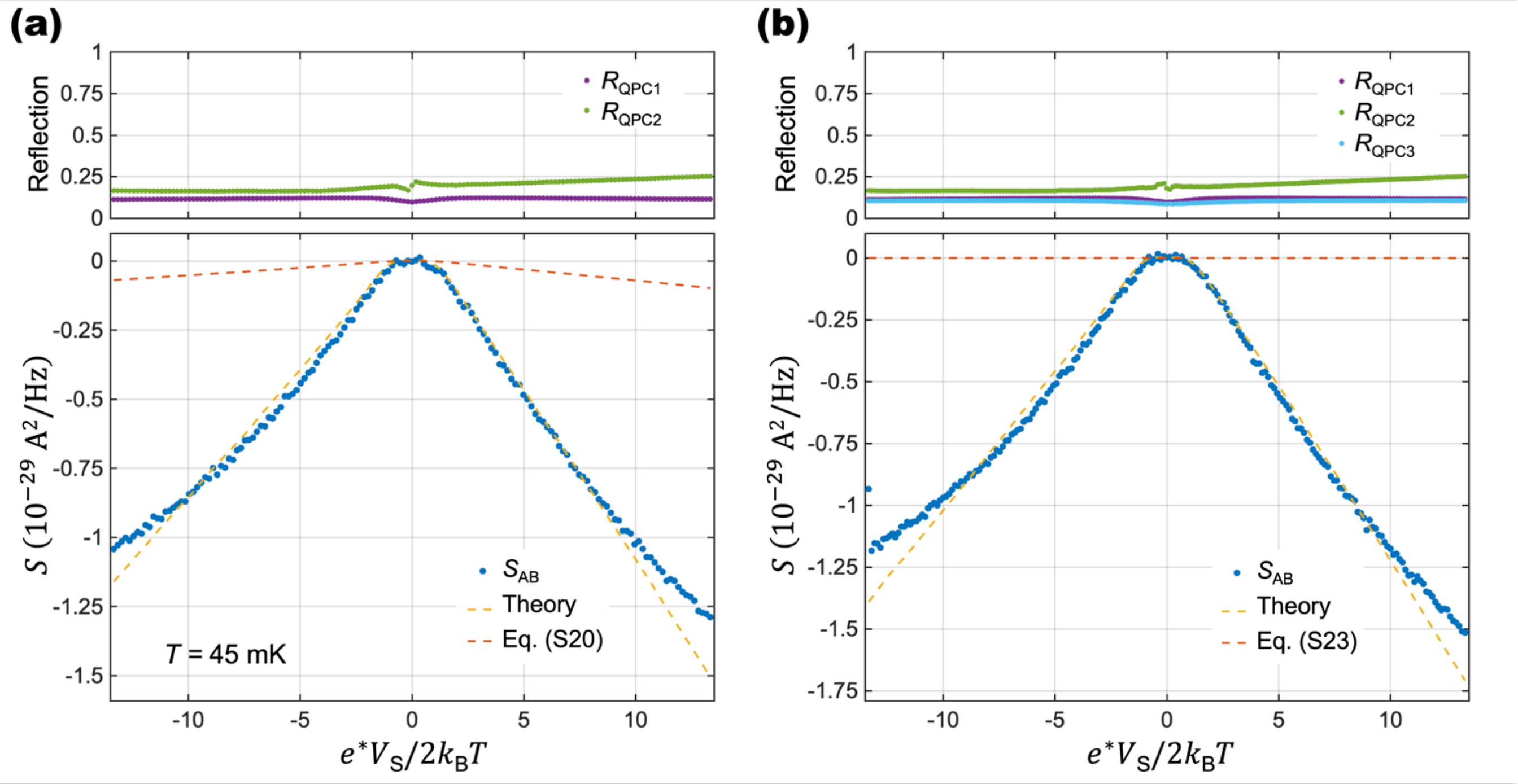

**FIG. S13: Cross correlations (a)** Cross correlation $S_{\mathrm{AB}}$ of the two-QPC geometry. **(b)** $S_{\mathrm{AB}}$ of the three-QPC geometry with symmetric injection of two dilute beams to QPC2. The experimental data of **(a)** were obtained with the same measurement (e.g., $R_{\mathrm{QPC1}}$, $R_{\mathrm{QPC2}}$) with Fig. 4(a) of the main text. The data of **(b)** were obtained with the average value of $R_{\mathrm{QPC1}} = 0.116$, $R_{\mathrm{QPC2}} = 0.192$, and $R_{\mathrm{QPC3}} = 0.102$. The results are in good agreement with our theoretical result (yellow dashed lines). The results of Eq. (S20) and Eq. (S23) for the trivial partition process (red dashed lines) are shown for comparison.

We also analyze our experimental data of the CC $S_{\mathrm{AB}}$ of a three-QPC geometry [Fig. S1(b)], which is essentially the same configuration with Ref. [S10]. To have the three-QPC geometry, we operated an additional QPC, QPC3, located downside of QPC2. This QPC connects Edge3 with an additional edge channel, Edge4, via anyon tunneling. By applying a voltage of the same amplitude $V_{\mathrm{S}}$ to the source contact S2 with that applied to the source contact S1 of Edge1, a current $I_{\mathrm{S}}$ flows along Edge4 (the same amount with the current along Edge1). It is reflected at QPC3, then a dilute beam of current $I_{\text{QPC3}} = R_{\text{QPC3}} I_{\text{S}}$ is generated to flow along Edge3 towards QPC2, where $R_{\text{QPC3}}$ is the reflection probability at QPC3. So the two dilute beams, one along Edge2 and the other along Edge3, are injected to QPC2. Then the CC $S_{\text{AB}}$ between amplifiers A and B was measured.

Theoretically, the CC is written as

$$S_{\text{AB}} = -S_{\text{QPC2}} + \frac{\partial I_{\text{QPC2}}}{\partial I_{\text{QPC1}}} S_{\text{QPC1}} - \frac{\partial I_{\text{QPC2}}}{\partial I_{\text{QPC3}}} S_{\text{QPC3}} , \tag{S21}$$

where $S_{\text{QPC3}} = 2e^* I_{\text{S}} R_{\text{QPC3}} (1 - R_{\text{QPC3}}) \left[\coth\left(\frac{e^* V_{\text{S}}}{2k_{\text{B}}T}\right) - \frac{2k_{\text{B}}T}{e^* V_{\text{S}}}\right]$ is the excess tunneling noise at QPC3 following Eq. (1). The third term is the correlation between the tunneling currents $I_{\text{QPC2}}$ and $I_{\text{QPC3}}$. As in our phenomenological theory for the two-QPC setup, we calculated the noise $S_{\text{QPC2}}$ as

$$S_{\text{QPC2}} = \mathcal{F}_{\text{dilute}} \times 2e^* |I_{\text{QPC1}} - I_{\text{QPC3}}| R_{\text{QPC2}} (1 - R_{\text{QPC2}}) \left[\coth\left(\frac{e^* V_{\text{S}}}{2k_{\text{B}}T}\right) - \frac{2k_{\text{B}}T}{e^* V_{\text{S}}}\right]. \tag{S22}$$

The Fano factor $\mathcal{F}_{\text{dilute}} = S_{\text{QPC2}}/2e^* I_{\text{QPC2}}$ is calculated by the CLL theory for $R_{\text{QPC2}} \ll 1$ and $e^* V_{\text{S}} \gg k_{\text{B}} T$ as before. The non-equilibrium correlator in Eq. (S4) has the multiplicative factor in the first term, which describes the effect of the dilute beam injected across QPC1. In the case of the two dilute beams, the non-equilibrium correlator is modified such that the first term has an additional multiplicative factor of $\left(1 - R_{\text{QPC3}} + R_{\text{QPC3}} e^{-2i\theta \text{sign}\,(t_2 - t_1)}\right)^{\frac{I_{\text{S}}}{e^*}|t_2 - t_1|}$ which describes the effect of the dilute beam injected across QPC3. We note that the braiding phase factor $e^{-2i\theta \text{sign}\,(t_2 - t_1)}$ of this multiplicative factor for the dilute beam injected across QPC3 differs from the factor $e^{2i\theta \text{sign}\,(t_2 - t_1)}$ of the multiplicative factor for the dilute beam injected across QPC1, because the time-domain loop at QPC2 braids the two beams in the opposite direction to each other. Using the modified non-equilibrium correlator, it is straightforward to compute the tunneling current and noise at QPC2, hence, the Fano factor $\mathcal{F}_{\text{dilute}}$.

Note that in the case of the perfectly symmetric injection of $I_{\text{QPC1}} = I_{\text{QPC3}}$, $\mathcal{F}_{\text{dilute}}$ diverges, and Eq. (S22) is invalid. However, Eq. (S22) is applicable to our experimental situation where there was about 10% difference between $R_{\text{QPC1}}$ and $R_{\text{QPC3}}$ [Fig. S13(b)] so that both $I_{\text{QPC1}} - I_{\text{QPC3}}$ and $\mathcal{F}_{\text{dilute}}$ are finite.

For comparison between our experimental data and the theory, we replace the differential reflections $\partial I_{\text{QPC2}}/\partial I_{\text{QPC1}}$ and $-\partial I_{\text{QPC2}}/\partial I_{\text{QPC3}}$ in Eq. (S21) by their averaged value $R_{\text{QPC2}}$. Then $S_{\text{AB}}$ is obtained by using $S_{\text{QPC2}}$ in Eq. (S22), $S_{\text{QPC1}}$ in Eq. (1) of the main text and an equation for $S_{\text{QPC3}}$ corresponding to Eq. (1). This theoretical result is in good agreement with our experimental data [Fig. S13(b)]. The excellent agreement between our phenomenological theory and our measurement of the CC $S_{\text{AB}}$ strongly supports that the time-domain braiding process is the underlying mechanism in both the two- and three-QPC geometries. Note that we also plot the non-interacting results from the Landauer-Büttiker formalism with the trivial partition process (with $e^* = e/3$),

$$S_{\text{AB}} = -2e^* I_{\text{S}} (R_{\text{QPC1}} - R_{\text{QPC3}})^2 R_{\text{QPC2}} (1 - R_{\text{QPC2}}) \left[\coth\left(\frac{e^* V_{\text{S}}}{2k_{\text{B}}T}\right) - \frac{2k_{\text{B}}T}{e^* V_{\text{S}}}\right], \tag{S23}$$

as the red dashed line in Fig. S13(b) for comparison.

Last but not least, we point out that measurement of AC $S_{\text{B}}$ at the port B in the two-QPC geometry is more useful for detecting the time-domain anyon braiding at QPC2 than the CC $S_{\text{AB}}$, especially for the case of non-Abelian anyons. It is firstly because $S_{\text{B}}$ is more directly related to the noise $S_{\text{QPC2}}$ at QPC2 where the time-domain braiding process happens. As shown in Eq. (S16), $S_{\text{B}}$ becomes the same with $S_{\text{QPC2}}$ as the temperature becomes lower. By contrast, the difference between the CC $S_{\text{AB}}$ and $S_{\text{QPC2}}$ is not negligible, as the second term of Eq. (S19) is of the same order with $S_{\text{QPC2}}$. Secondly, in the most promising non-Abelian FQH states, upstream and downstream flows coexist along FQH edges. Then, there can occur some side-effects by the coexistence [S1]. The ratio of the side-effects compared to the main signal of our interest is of the order of $R_{\text{QPC1}} R_{\text{QPC2}}$ for the case of $S_{\text{B}}$, but it is of the order of $R_{\text{QPC1}}$ for the case of $S_{\text{AB}}$. The AC $S_{\text{B}}$ is more robust against the side-effects than CC $S_{\text{AB}}$.